\documentclass[3p]{elsarticle}

\usepackage{amssymb}
\usepackage{amsmath}

\usepackage{hyperref}
\usepackage{algorithm}
\usepackage{algpseudocode}
\makeatletter
\newenvironment{model}[1][htb]
  {\renewcommand{\ALG@name}{Model}
   \begin{algorithm}[#1]%
  }{\end{algorithm}}
\makeatother

\usepackage{enumitem}
\newlist{questions}{enumerate}{2}
\setlist[questions,1]{label=RQ\arabic*.,ref=RQ\arabic*}
\setlist[questions,2]{label=(\alph*),ref=\thequestionsi(\alph*)}

\usepackage{graphicx}
\usepackage{svg}
\usepackage{color}
\usepackage{caption}
\usepackage{subcaption}
\usepackage{natbib}
\usepackage{booktabs} 

\usepackage{soul}

\usepackage{verbatim}

\usepackage{xargs}     
\usepackage[colorinlistoftodos,prependcaption,textsize=tiny]{todonotes}
\newcommandx{\unsure}[2][1=]{\todo[linecolor=purple,backgroundcolor=purple!25,bordercolor=purple,#1]{#2}}
\newcommandx{\changed}[2][1=]{\todo[linecolor=blue,backgroundcolor=blue!25,bordercolor=blue,#1]{#2}}
\newcommandx{\added}[2][1=]{\todo[linecolor=green,backgroundcolor=green!25,bordercolor=green,#1]{#2}}
\newcommandx{\delete}[2][1=]{\todo[linecolor=red,backgroundcolor=red!25,bordercolor=red,#1]{#2}}

\usepackage{eurosym}  
\newcommand{\eurpkwh}{€ per kWh}
\DeclareUnicodeCharacter{2212}{-} 
\DeclareUnicodeCharacter{2265}{$\geq$} 
\usepackage{lmodern,textcomp}

\usepackage{nomencl}
\makenomenclature

\renewcommand{\nomgroup}[1]{%
  \ifthenelse{\equal{#1}{S}}{\item[\textbf{Subscripts and Sets}]}{%
  \ifthenelse{\equal{#1}{P}}{\item[\textbf{Parameters}]}{%
  \ifthenelse{\equal{#1}{V}}{\item[\textbf{Variables}]}{%
  \ifthenelse{\equal{#1}{A}}{\item[\textbf{Abbreviations}]}{}}}}}

\journal{Applied Energy (Elsevier)}

\begin{document}
\newpage

\begin{frontmatter}
\title{Optimal Sizing and Control of a Grid-Connected Battery in a Stacked Revenue Model Including an Energy Community}

\author[1,2,6]{Tudor Octavian Pocola}
\author[1,3,6]{Valentin Robu}
\author[4]{Jip Rietveld}
\author[5]{Sonam Norbu}
\author[5]{Benoit Couraud}
\author[5]{Merlinda Andoni}
\author[5]{David Flynn}
\author[6]{H. Vincent Poor}
\address[1]{Intelligent and Autonomous Systems Group, CWI (National Research Centre for Mathematics and Computer Science), Amsterdam, The Netherlands}
\address[2]{Algorithmics Group, Delft University of Technology, Delft, The Netherlands}
\address[3]{Electrical Engineering Department, Eindhoven University of Technology, Eindhoven, The Netherlands}
\address[4]{GIGA Storage BV, Amstelveen, The Netherlands}
\address[5]{James Watt School of Engineering, University of Glasgow, Glasgow, UK}
\address[6]{Electrical and Computer Engineering, Princeton University, Princeton, NJ, USA}

\begin{abstract}
Recent years have seen rapid increases in intermittent renewable generation, requiring novel battery energy storage systems (BESS) solutions. One recent trend is the emergence of large grid-connected batteries, that can be controlled to provide multiple storage and flexibility services, using a stacked revenue model. Another emerging development is renewable energy communities (REC), in which prosumers invest in their own renewable generation capacity, but also requiring battery storage for flexibility. In this paper, we study settings in which energy communities rent battery capacity from a battery operator through a battery-as-a-service (BaaS) model. We present a methodology for determining the sizing and pricing of battery capacity that can be rented, such that it provides economic benefits to both the community and the battery operator that participates in the energy market. We examine how sizes and prices vary across a number of different scenarios for different types of tariffs (flat, dynamic) and competing energy market uses. Second, we conduct a systematic study of linear optimization models for battery control when deployed to provide flexibility to energy communities. We show that existing approaches for battery control with daily time windows have a number of important limitations in practical deployments, and we propose a number of regularization functions in the optimization to address them. Finally, we investigate the proposed method using real generation, demand, tariffs, and battery data, based on a practical case study from a large smart battery operator in the Netherlands. For the settings used in our case study, we find that a community of 200 houses equipped with a 330 kW wind turbine can save up to €12,874 per year by renting just 280 kWh of battery capacity (after subtracting the battery rental costs), and the methodology is applicable for a wide variety of other settings and tariff types.
\end{abstract}



\begin{keyword}
Battery as a Service\sep Smart Battery Control\sep Energy Communities\sep Optimization
\end{keyword}

\end{frontmatter}


\renewcommand{\nomname}{}


\nomenclature[P]{$g_i^{\text{wind}}$}{Wind power generation at time step $i$}
\nomenclature[P]{$d_i$}{Community demand at time step $i$}
\nomenclature[P]{$\tau^b_i$}{Grid buying price at time step $i$}  
\nomenclature[P]{$\tau^s_i$}{Grid selling price at time step $i$}
\nomenclature[P]{$\tau^{DA}_i$}{Day-ahead market prices at time step $i$}  
\nomenclature[P]{$\eta_c$}{Battery charging efficiency}  
\nomenclature[P]{$\eta_d$}{Battery discharging efficiency}  
\nomenclature[P]{$\eta_{cd}$}{Battery round-trip efficiency}  
\nomenclature[P]{$SoC^{\text{initial}}$}{Initial state of charge of the battery} 
\nomenclature[P]{$SoC^{\text{max}}$}{Maximum state of charge of the battery}  
\nomenclature[P]{$SoC^{\text{min}}$}{Minimum state of charge of the battery}  
\nomenclature[P]{$SoC^{EoD}$}{Minimum state of charge at the end of day}  
\nomenclature[P]{$p^{\text{max}}$}{Maximum charging and discharging power of the battery} 
\nomenclature[P]{$e^{\text{max}}$}{Maximum import and export energy}  

\nomenclature[P]{$\Delta t$}{Time step duration}  
\nomenclature[P]{$\lambda^{\text{charging}}$}{Penalty for battery charging and discharging}  
\nomenclature[P]{$\lambda^{\text{capacity}}$}{Penalty for an empty battery at the end of the day}  
\nomenclature[P]{$\lambda^{\text{max\_cycles}}$}{Maximum allowed battery cycles per day}

\nomenclature[V]{$SoC_i$}{Battery state of charge at time step $i$}  
\nomenclature[V]{$p^{\text{charge}}_i$}{Battery charging power at time step $i$} 
\nomenclature[V]{$p^{\text{discharge}}_i$}{Battery discharging power at time step $i$}  
\nomenclature[V]{$e^s_i$}{Energy exported to the grid at time step $i$}  
\nomenclature[V]{$e^b_i$}{Energy imported from the grid at time step $i$}

\nomenclature[A]{BaaS}{Battery as a Service}
\nomenclature[A]{BESS}{Battery Energy Storage System}
\nomenclature[A]{EoD}{End of Day}
\nomenclature[A]{LP}{Linear Programming}
\nomenclature[A]{MILP}{Mixed Integer Linear Programming}
\nomenclature[A]{PV}{Photovoltaic}
\nomenclature[A]{REC}{Renewable Energy Community}
\nomenclature[A]{SoC}{State of Charge}
\nomenclature[A]{TSO}{Transmission System Operator}
\nomenclature[A]{TSO}{Transmission System Operator}
\printnomenclature

\section{\label{cha:intro}Introduction}

In recent years, there has been a significant increase in the adoption of renewable energy sources such as solar and wind power. According to the International Energy Agency (IEA), global renewable electricity capacity additions increased by nearly 50\% in 2023, reaching 507 gigawatts (GW), and setting a new record for the 22nd consecutive year~\cite{IEA2023}.   
Yet, the integration of renewable energy sources into the power grid requires substantial investment in 
~energy storage solutions~\cite{MITALI2022166} to manage the intermittent nature of renewable generation and to ensure a reliable and constant energy supply~\cite{Seia2Nov023,owid-renewable-energy}.

Among storage options, lithium-ion Battery Energy Storage Systems (BESSs) have seen increased commercial deployment in recent years, due to significant cost reduction \cite{ORANGI2024109800} and desired technical properties of high power throughput, high round-trip efficiency and fast response times. The Solar Energy Industry Association (SEIA) has forecast a sixfold increase in demand for BESS in the US and a sector growth reaching to 840 GWh by 2030 from 60 GWh in 2022 worldwide~\cite{Seia2Nov023}. Commercial deployments include
~large-scale, grid-connected batteries. Examples include (among many others): the 300 MW / 450 MWh large battery project in Victoria, Australia\footnote{\url{https://victorianbigbattery.com.au/}} (one of the first utility-scale projects, using Tesla mega-pack batteries), the 875 MW / 3,287 MWh Edwards \& Sanborn Solar Plus Storage Project\footnote{\url{https://www.power-technology.com/projects/edwards-sanborn-solar-and-energy-storage-project-california-usa}}
~in California, USA, and the 98MW / 196 MWh system in Yorkshire, UK, currently being developed in conjunction with the Dogger Bank offshore wind park\footnote{\url{https://www.datacenterdynamics.com/en/news/uk-builds-europes-biggest-battery-out-of-tesla-megapacks/}}. In the Netherlands, large battery projects include the Rhino and Buffalo batteries (12 MW / 7.5 MWh and 25 MW / 48 MWh, respectively) operated by GIGA Storage\footnote{\url{https://giga-storage.com/projecten/}}, the project partner who also provided the case study used to test models developed in this paper. 


These
~large-scale, grid-connected batteries are usually independently owned and operated for commercial purposes.
~Yet, because of their high initial investment and replacement costs arising from degradation,
~grid-connected batteries
~have to participate in several end-uses and markets, stacking several revenues to reduce the payback period and improve their financial viability. The concept of {\bf revenue stacking} is an emergent business model that optimizes battery capacity by allocating it across multiple uses and market applications. While they are often deployed and located with the aim of balancing local renewable generation at a specific site (e.g. large solar park or wind farm), to improve their business case, batteries may be operated to also provide other services, including participation in wholesale markets (day-ahead and spot markets)~\cite{Tennet-Markets, Hartman2022} and ancillary services. Various studies have highlighted the importance of BESS in the provision of grid services, such as voltage support, frequency regulation, black-start, congestion relief, peak-shaving, and power smoothing~\cite{Al-Foraih2018Sep, Castillo2023-Ancilaryservices, MirMohammadiKooshknow2018Dec,  Prakash2022Sep, su14105985,  ZAKERI2015569}. However, the key challenge in revenue stacking is determining the optimal times for charging and discharging the battery in response to shifting market prices, while also minimizing battery degradation \cite{Alharbi}.

Another recent development in the fast-evolving energy landscape is the emergence of energy communities, such as the Renewable and Citizen Energy Communities (REC and CEC) introduced by EU legislation~\cite{EUdirective1, EUdirective2}. These communities consist of multiple prosumers \cite{Parra-Dominguez2023Jul} (i.e. households or commercial entities that both produce and consume energy), who combine their energy assets to optimize their energy use, reduce costs, and enhance the consumption of locally sourced renewable energy. Energy communities represent a decentralized approach to energy management~\cite{Guedes2022Oct}, enabling individual prosumers to control their own energy supply, trade energy with each other in a peer-to-peer (P2P) way, or jointly invest in energy generation assets~\cite{sousa2019peer,tushar2019motivational,capper2021systematic,norbu_constraints}. 

Batteries ranging from single-household deployments (e.g., Tesla Powerwall) to larger-scale community systems~\cite{norbu_constraints, Norbu2021Apr,couraud_ReFLEX} can play a crucial role in energy communities, offering local balancing and enhancing reliability. However, community members face significant challenges when installing battery systems due to their large upfront cost and lack of technical know-how regarding their operation, especially in the context of revenue stacking models. For example, grid-connected batteries often require costly certification processes in order to allow them to be used for secondary revenue streams, such as from trading in the day-ahead/balancing markets, or providing grid services. Prosumers lack the expertise to participate in such more advanced, but highly profitable markets, thus they can miss out on important revenue streams for their battery assets. In this work, we take a different perspective, which often appears in practical settings: the operator of a large, grid-connected battery, that has the required size and expertise to participate in multiple markets, can integrate an energy community into their revenue stacking model by leasing part of their battery capacity.
~The revenue obtained from energy communities, while perhaps less profitable than from trading energy in the power markets, is often less volatile and more predictable - leading to a more sustainable business model. However, integrating them into a stacked revenue model for grid-scale batteries raises key knowledge gaps, including:
~\emph{How much battery should be hired for an energy community's use vs. other alternative markets?  What prices can be charged (for different capacity levels) to make the proposition economically attractive for both the battery operator and the energy community in question? How is this affected by different levels of import/export tariffs that are available to prosumers on the local market?}  Currently, to the best of the authors' knowledge, no rigorous, data-driven methodology exists for answering such questions - which can then be tailored for specific cases, market scenarios, and energy communities.

Another important challenge for operators of large grid-connected batteries is the question of optimal battery control. While the literature on battery control has grown considerably in recent years, many existing approaches - both in academic literature and in practice - apply linear programming to optimize over day-ahead (24-hour) time windows. This is a natural approach, as with modern forecasting techniques, the mean expected renewable generation (which depends on expected solar irradiance and wind speed at each location), can provide relatively good accuracy over day-ahead (24-hour) windows. However, a straightforward application of this optimization approach can have a number of shortcomings that considerably reduce its performance in practical applications. First, it leads to a potentially very large number of battery charge/discharge cycles each day, which would not be implemented in practice due to potential battery state-of-health degradation considerations. Second, it can lead to excessive discharging of the battery at the end of individual time windows, which is a lost opportunity for battery operators. Hence, we need some ways to guide linear optimization
~approaches towards solutions that do not exhibit these issues, and we show how this can be achieved through adding regularization constraints in the linear programming optimization.

Finally, in order for any developed methodologies to be accepted and used in practice, their performance needs to be experimentally verified in a wide range of settings, using real datasets, for different energy market setups, import/export tariffs, and renewable energy capacities. In this paper, we start from the real-world case study of one of the largest operators of grid-connected batteries in the Netherlands, GIGA Storage in Amsterdam, and perform a full examination of the robustness of our methods (along various dimensions) using data from the Netherlands and the UK. 
While the model developed in this work played an important practical role in allowing GIGA to evaluate the potential of incorporating communities in their stacked revenue model, the methodology proposed is much broader and can be used in a variety of settings and communities worldwide to tackle this emerging challenge. 
In summary, our study makes the following contributions to the state of the art:

\begin{itemize}
\item First, we propose a principled methodology for allocating 
battery capacity within a stacked revenue model, integrating its use in both a renewable energy community and the energy day-ahead market. More specifically, our methodology allows the battery operator 
to determine the battery capacity that is economically feasible to rent to communities, based on determining minimum and maximum price ranges that can be accepted for both parties. This methodology allows battery operators and communities to find mutually beneficial battery sizing and pricing agreements.

\item Second, we perform a thorough investigation of models for the control of grid-connected batteries. We particularly focus on two practical challenges we identify when implementing linear optimization methods, namely that, first, optimization with fixed time windows (typically day-ahead windows) can return solutions with too many battery charge/discharge cycles and, second, can lead~to excessive discharge of the battery at the end of each time window. 
We propose two regularization factors that overcome these problems, providing efficient, practical methods for battery control.

\item Finally, we comprehensively test our proposed methodology and control algorithms for a variety of market setups and types of tariffs, using real-world data for generation, demand, and market prices from Western Europe (i.e. Netherlands and the UK). In more detail, our evaluation is based on the case study of GIGA Storage, one of the largest operators of grid-connected batteries in the Netherlands.
\end{itemize} 

This paper is organized as follows.
~Section ~\ref{cha:related} provides a thorough literature review of related work of key concepts of this work,
~Section ~\ref{cha:methodology} presents our methodology and technical contributions, while Section ~\ref{cha:exp} discusses the results from the different data-driven experiments used to verify the model. Section ~\ref{cha:disc} provides a discussion of our results, exploring their implications. Finally, Section ~\ref{cha:conc} concludes by briefly summarizing the main findings, and outlines some directions for future work.

\section{\label{cha:related}{Related Work}}

This section presents a comprehensive review of the literature organized into three main sections aligned with the research questions: Battery Energy Storage Systems (BESS) as a service, Renewable Energy Communities (REC), and Optimal Battery Control. The first section explores the various applications of BESS identified in the literature with focus on the identification of potential revenue streams that could be used in a
~stacked revenue model.
~The second section examines the emerging trend of RECs, discussing the critical role of BESS in these communities and the configurations under which they are implemented especially regarding
~techno-economic analysis and policy-making aspects. The final section focuses on general control methods for BESS, discussing technical perspectives and control strategies, and examining studies that address these issues from a more theoretical standpoint. By clustering the literature in this manner, this review aims to provide a structured understanding of the current state of research and identify gaps that this paper seeks to address.




\subsection{BESS as a service}
This section reviews research work around BESS with a focus on the  identification of the most promising battery use cases in electricity markets, and studies that have implemented revenue stacking models including in energy communities.

Gulotta et al. \cite{Gulotta2023Mar} focused on the integration of distributed energy resources (DERs), including batteries, into ancillary services. They highlight the important role of BESS in the transition to decentralized systems, especially its economic benefits and its contribution to grid reliability. Furthermore, the study emphasizes the need for improved regulatory frameworks to facilitate the participation of BESS in the ancillary services markets. Prakash et al. also reviewed BESSs for ancillary services and highlighted three critical applications: frequency regulation (FR), voltage support, and peak shaving~\cite{Prakash2022Sep}. In the same vein, Castillo et al. \cite{Castillo2023-Ancilaryservices} identified BESS application for ancillary services, including frequency regulation, voltage support, arbitrage, peak shaving, load smoothing, and black start. The work focused on control strategies, technical challenges and market rules, while it emphasized the importance of optimizing the battery sizing and location to maximize economic benefits and enhance grid reliability. Kooshknow and Davis \cite{MirMohammadiKooshknow2018Dec} mapped profitable single use cases for the Dutch market covering multiple actors across the electricity supply chain. They report the reserve market as the main application, but anticipate an increasing demand for flexibility services.

While these studies have identified potential revenue streams from single use cases, the importance of integrating multiple multiple revenues in the battery operation and control has also been highlighted in the literature. For example, Marnell et al. \cite{Marnell2019} reviewed transmission-scale battery technologies with a focus on potential revenue streams and identified the provision of ancillary services (voltage support, black start, reserves), arbitrage, congestion relief and islanding support, as potential revenue streams. In addition, they also discussed the significance of a stacked revenue model, where BESS scheduling is achieved by a joint optimization function across multiple services. Biggins et al. \cite{BIGGINS2022104234} implemented a stacked revenue model combining price arbitrage and participation in the FR market. They first predict the likelihood of their battery being accepted for FR services, then incorporate this into a mixed-integer linear programming (MILP) optimization model for optimal battery scheduling. While their results show that FR is the largest source of revenue, arbitrage was also shown to be feasible within a small, risk-constrained band in real-time operations. By analyzing several business scenarios for 10 battery installations in Finland, Ramos et al \cite{RAMOS2021102720} evaluated Battery as a Service (BaaS) models, including service provision to end users or communities. Through stakeholder interviews, they identified gaps between theoretical models and practical implementations, and emphasized the importance of revenue stacking to improve BESS cost effectiveness and adoption in energy community projects.

In summary, the literature review reveals that batteries have promising use cases both in the wholesale and ancillary services markets, and that simultaneous service provision across markets may be more profitable. On the other hand, BaaS models that also integrate energy communities into a revenue stacking model, have not yet received sufficient research interest, indicating a gap in the existing literature. The practical case study that inspired this work, aligns well with these findings. Based on the experience of our commercial partner GIGA Storage, the day-ahead market and the imbalance market currently provide the most profitable avenue for use of large, grid-connected batteries in the Netherlands, although ancillary markets are a commonly reported application for BESS in the literature. This paper aims to fill this gap by exploring the feasibility and potential benefits of implementing BaaS within energy communities, while also participating in the day-ahead market. More specifically, we focus our attention on the revenues from the day-ahead market, comparing it to potential revenues from the local community. The reason for this the predictability: while revenues on the imbalance market can be potentially higher, they are highly volatile. This means they are highly dependent on having good price forecasts and carry considerable market risks. Revenues in the day-ahead market (while not risk-free) are more predictable and can be reliably modeled with the use of linear programming optimization, thus representing a better benchmark to compare our community approach, in a stacked revenue model. The following section reviews the role of batteries in energy communities to increase understanding of the context of the paper.

\subsection{Energy Communities and the role of BESS}
Recent years have seen significant changes in the regulatory context aiming at the empowerment of end users and their active engagement in the energy transition. Notable examples are the legislative efforts of the EU, namely the Renewable and Citizen Energy Communities (REC and CEC) introduced in EU directives~\cite{EUdirective1, EUdirective2}. These initiatives aim to enable end users to collectively engage in energy production, management and sharing, including decentralized models of P2P energy trading.

Several research studies have focused on the topic of energy communities. Ahmed et al. \cite{Ahmed2024-review} offers a comprehensive review of RECs, exploring their concepts and benefits in enhancing energy independence and resilience.  The authors highlight the importance of energy storage, such as batteries, for communities, however, challenges like high initial costs and potential technical issues with battery maintenance and lifespan are noted as disadvantages. Despite these drawbacks, batteries remain promising solutions 
~\cite{CHEN2009291}, emerging as the most cost-effective in the current economic climate, although they have a higher environmental impact compared to other technologies \cite{HANNAN2021103023}. Belmar et al. \cite{belmar2023modelling} model different design configurations and local energy market setups for energy communities in Lisbon, Portugal. They found the highest cost savings were reached when prosumers were equipped with both PV and battery systems. Guedes et al. \cite{Guedes2022Oct} incorporated BESS into energy communities of prosumers with generation and storage capabilities. They implemented a linear optimization model aimed at increasing the community income, while taking battery degradation into account, and evaluated the results using social welfare and fairness indicators. It was shown that batteries improved community satisfaction while also increasing the social welfare of the market. Davis and Hiralal \cite{DAVIS20161448} evaluated the economic feasibility of UK prosumers investing in individual BESS to flatten their electricity consumption or shift daytime demand to cheaper nighttime rates. Both scenarios proved unfeasible without government subsidies due to high upfront costs. Most importantly, the authors discussed BaaS as a viable alternative to community ownership. Pasqui et al. \cite{PASQUI2023} highlighted the importance of government incentives and third-party providers in ensuring the economic feasibility and operational efficiency of RECs. Their analysis of an Italy-based REC compared four battery management strategies: no battery, standard, smart, and optimal. While the optimal strategy maximized community profits, its reliance on perfect foresight of generation and consumption made it impractical. Instead, the smart approach proved more viable, enhancing self-consumption with only a minor drawback for prosumers. Aranzabal et al. \cite{Aranzabal2023Jan} proposed an advanced three-level control strategy for BESS scheduling that leads to revenue maximisation of a REC. In their model the community is treated as a virtual microgrid and participates in the automatic frequency restoration reserves (aFRR) ancillary market. The control strategy incorporated load and generation forecasting, MILP optimization for battery scheduling and use of decision-trees for real-time adjustments and demonstrated its viability in several use cases and scenarios.

The literature review reveals the importance of batteries in energy communities, however the use of BaaS and revenue stacking models is limited. This work proposes a strategy for allocating battery capacity within a stacked revenue model, integrating its use in both a renewable energy community and the energy spot market. However, to effectively engage with the wholesale and ancillary markets with a stacked revenue model, BESS owners must deploy competitive and efficient control algorithms. Optimal battery control strategies are reviewed in the following section.

\subsection{\label{sec:battery_control}Optimal Battery Control}
A large number of research papers can be found in the literature on optimal battery control strategies. In this work, we focus our attention in LP approaches used for optimal battery scheduling and report the advantages and limitiations of such methods. There has been an increase in recent years in modeling BESS as a linear set of constraints \cite{Sioshansi-9513574}. These models provide a well-understood methodology for analyzing the optimal behavior of BESS across various problems. However, LP models lack an explicit formulation that accounts for mutually exclusive operations, such as charging and discharging. Additional constraints can be added to address this issue, but have been shown (c.f. Jose et al.~\cite{Jose-9183968}) to significantly slow down the model. This occurs because the added constraints contain binary variables, converting the model into a MILP model, which is considerably slower. Zhao et al. \cite{Zhao-8396307} suggested that the binary representation, which can slow down the computation, can be omitted if the battery's round-trip efficiency is less than one.
A more recent study by Pozo \cite{POZO2022108565} directly addresses this issue by analyzing key linear BESS models and their limitations. The study examines one MILP model and four LP models in the context of two classical power system problems: set-point tracking and transmission expansion planning. Among these, a simplified LP model is widely used in techno-economic analysis, while another approach ensures feasible charging schedules by respecting state-of-energy limits. A relaxed LP formulation improves computational efficiency but tends to overestimate the objective function, sometimes by as much as 15\%. Pozo also introduces a model that closely aligns with the MILP formulation. The study finds that only the MILP-based approach fully avoids simultaneous charging and discharging, though at the cost of significantly higher computational time. While differences among LP models are notable in short-term tracking problems, they become negligible for long-term planning decisions.
Qin et al. (2016) \cite{Qin-IEEE} formulate the problem of optimal energy storage operation under uncertainty as a stochastic control problem with general cost functions. They propose a simple yet effective heuristic algorithm and demonstrate that, while its control is not fully optimal, its deviation from the optimal solution is mathematically bounded by system parameters. This bound is easily computable, making the approach both practical and theoretically sound for evaluating other heuristic methods. They further extend the algorithm to a distributed setting \cite{Qin2-IEEE}, enabling networked storage operation under uncertainty, which improves scalability and applicability in larger, more complex systems.

As found in the literature, there is a trade-off between the speed of convergence and accuracy of the optimization methods. As the research works above, this work also focuses on practical challenges experienced when implementing LP optimization methods, namely that, first optimization with fixed time windows (typically day-ahead windows) can return solutions with too many battery charge/discharge cycles and, second, can lead to excessive discharge of the battery at the end of each time window. To solve this issues, we propose two regularization factors that
overcome these problems, providing efficient, practical methods for battery control. The methodology applied is presented in detail in the following section. 
\section{\label{cha:methodology}Methodology}
\begin{figure}[t]
    \centering
    
    \includegraphics[width=1.\textwidth]{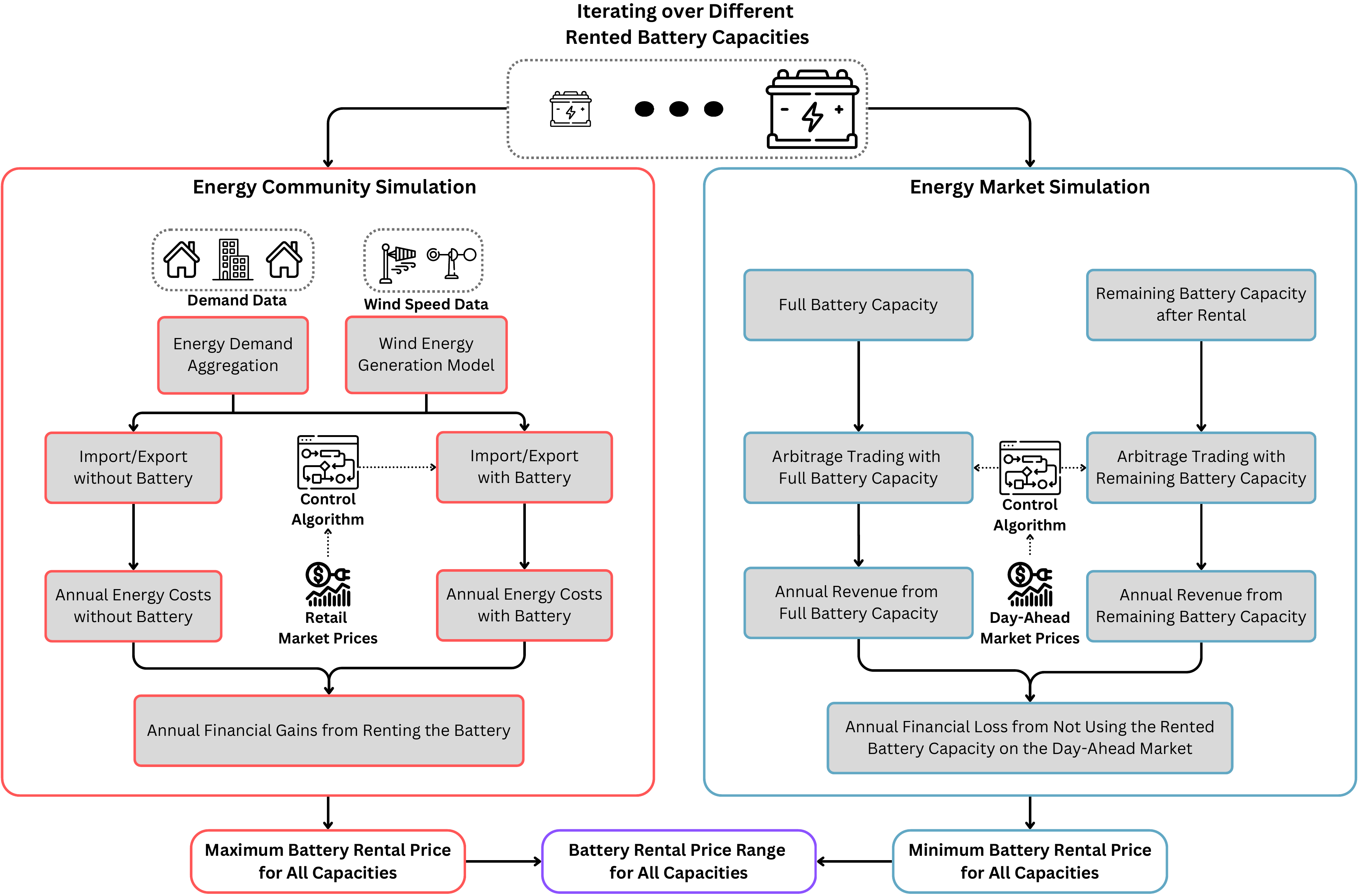}
\caption{\textbf{Overall methodology for allocation of battery capacity between energy community and energy market use.} The left side of the diagram shows the calculation of potential savings from battery usage in the community, which informs the maximum rental price it is willing to pay. The right side shows  the calculation of expected profits from battery participation in the day-ahead market. The final rental price is determined within a range: the upper bound is the maximum amount the community is willing to pay based on their savings, while the lower bound is set by the minimum price required to remain competitive with market trading profits.}

    \label{fig:graphical_methodology}
\end{figure}

This section outlines the methodology used to assess the feasibility of allocating a portion of a grid-connected Battery Energy Storage System (BESS) to a Renewable Energy Community (REC) while maintaining its participation in energy markets. The objective is to determine an optimal rental price for the shared battery capacity, ensuring financial benefits for both the REC and the battery owner. Our revenue-sharing framework, illustrated in Figure~\ref{fig:graphical_methodology}, is divided into two main components: battery use within the energy community and battery use in the energy markets.

We begin by detailing how real-world data from the Netherlands and the United Kingdom was sourced to provide the necessary inputs for our simulations. Next, we introduce the control algorithms employed in this study. Using these elements, we derive feasible price ranges that ensure profitability for both parties.

From the REC’s perspective, we calculate the maximum rental price, which represents the highest price the community is willing to pay based on its potential savings under various tariff schemes. From the battery owner's perspective, we determine the minimum rental price, which corresponds to the lowest price they would accept, reflecting the profits the battery could generate if the same capacity were allocated to external energy markets, specifically the day-ahead market. By analyzing both perspectives, we establish a feasible price range that ensures financial viability for both stakeholders. In order to determine the optimal split between these revenue streams, we conduct multiple yearly simulations with varying battery capacity allocations, evaluating potential earnings across different scenarios to identify the configuration that maximizes overall profitability.

A key focus of this study is also to evaluate control strategies for managing the battery within the REC. Two approaches are compared: a \emph{daily linear programming (LP) optimization}, which uses a one-day planning horizon to optimize costs, and a \emph{heuristic-based greedy algorithm}, which operates without forecasts by reacting to the current energy state. These methods are assessed in terms of their effectiveness in maximizing savings and profits. 

Finally, we enhance the existing daily LP approach by introducing two regularization costs. These functions help guide the optimizer toward a more balanced and efficient yearly schedule, reducing excessive battery cycling while improving long-term profitability. We test our proposed method against different benchmarks, which attempt to achieve the same behavior. The following sections discuss each methodology component in detail.

\subsection{Modelling Energy Communities and Battery Usage}
\label{sec:meth:community_params}

To simulate battery operation within an energy community, we use two key data inputs: the \emph{energy demand}, based on consumption profiles of households and the \emph{generation curve} of renewable energy sources (e.g., solar or wind). These data sets are used to model energy flows within the community and serve as inputs for the control algorithms, which determine the battery’s \emph{state of charge (SoC)} over time.

\subsubsection{Community Demand \label{sec:demand_data}}

In this study, real energy demand data was sourced from the Thames Valley Vision (TVV) dataset \cite{TVV1}. The dataset includes demand curves for 200 class-1 and 20 class-2 households from the UK \cite{TVV-Data}. Specifically, for our study, we use the demand data from 200 class-1 households (following~\cite{Norbu2021Apr}), which represent unrestricted domestic consumers (note that, here, ``unrestricted" consumers denote those that are not given special tariff incentives to shift their demand profiles, as such incentives could affect their consumption patterns, and hence the results). The total yearly demand for the community is approx. 840.34 MWh, with an average yearly demand per household of 4201.71 kWh. Household demand varies significantly, with the smallest household consuming 1000 kWh annually and the largest consuming 18,735.47 kWh.

The demand data is provided in half-hourly intervals, meaning each household’s demand is recorded every 30 minutes. The demand for household \( j \) at time \( i \) is denoted as \( d_i^j \). To obtain the total demand for the community at each time step, we sum over all households, as follows:

\[
d_i = \sum_j d_i^j
\]

This total community demand \( d_i \) will serve as a key input for the battery control algorithms.


\subsubsection{\label{sec:wind_generation}Estimation of Community Wind Energy Generation}

The estimation of wind energy generation follows a methodology previously applied by Fruh \cite{Fruh2013Feb}, Andoni et al. \cite{Andoni2017Sep}, and Norbu et al.~\cite{Norbu2021Apr}. We include in this paper for completeness and reproducibility, but leave the full presentation of the method to \ref{appendix:wind_generation}. This approach uses real wind speed data to estimate the energy output of a wind turbine based on wind speed at a specific hub height.

For this study, meteorological data from the Royal Netherlands Meteorological Institute (KNMI) was employed\footnote{The wind speeds were retrieved from the \href{https://www.tudelft.nl/en/ewi/over-de-faculteit/afdelingen/electrical-sustainable-energy/photovoltaic-materials-and-devices/dutch-pv-portal/meteorological-data}{TU Delft Meteorological Portal}.}, focusing on data from North Holland. The wind speeds represent hourly averages measured at a height of 10 meters. These measurements were then adjusted to the turbine hub height using standard wind profile equations. The wind speeds at hub height were converted to energy output using the power curve of the Enercon E-33 wind turbine, which was approximated with a sigmoid function. The energy generation at time \( i \) is denoted as \( g_i^{\text{wind}} \).

To align the wind generation data with the half-hourly demand data, missing values in the wind speed dataset were handled using double-spline interpolation. This ensures that wind data matches the half-hourly resolution of the demand dataset, enabling a consistent input for the control algorithms. The cost of the wind turbine was incorporated into the calculations by amortizing it over a 20-year period, ensuring a realistic assessment of the financial implications of wind energy generation.

An important consideration is how much renewable energy generation a community should install -- especially in relation to local demand. For this, we formally define a \emph{wind generation coefficient} for the installed renewable generation as $R_{gen}=\frac{\sum_i^T{g_{i}}}{\sum_i^T{d_{i}}}$, where $\sum_i^T{g_{i}}$ is the total generation over all the timepoints in period $T$ (in this paper, all the half-hours in a year) and $\sum_i^T{d_{i}}$ is the total demand over the same period.
In the case of this paper, this coefficient refers to the scale of the wind turbine capacity, but it can also be, e.g., the size of PV panels, for communities with local solar generation. The coefficient of the original wind turbine used in this study (Enercon E-33 wind turbine~\cite{EnerconE33}) is \( R_{gen} = 1.2 \). However, we have examined other scaling coefficients in our experiments, particularly in the co-optimization of local renewable generation capacity and battery size, as presented in Section~\ref{sect:renewable_sizing_optimization}. A full description of the procedure to generate renewable generation (wind) data, including the handling of missing data, wind speed interpolation, and the turbine power curve estimation, can be found in \ref{appendix:wind_generation}.

\subsection{Modelling the Wholesale \& Retail Energy Markets}

\label{sec:meth:energy_prices}
Another required input consists of the wholesale and retail energy prices. These are needed not only to calculate the financial outcomes of importing and exporting energy but also to inform algorithms that optimize operations by looking ahead. From the perspective of the wholesale market, this study focuses on day-ahead market prices. From the perspective of the retail market, we consider two types of tariffs: fixed and dynamic.

\subsubsection{Day-Ahead Energy Markets}

From the perspective of large battery operators, such as our case study partner GIGA Storage, participation in the day-ahead market represents a key revenue source.
Market prices for the year 2023 were obtained from ENTSO-E\footnote{\url{https://www.entsoe.eu/}}, using an open-source package \cite{Pecinovsky_entsoe-py}. Normally, market prices include a transport tariff. However, in the context of the Netherlands (based on the GIGA Storage case study), large grid-connected battery operators that have a contract with the TSO can import energy without paying the transport costs. 

The day-ahead spot trading market is particularly relevant for grid-connected batteries in the Netherlands. Linear optimization has been shown to work with a high degree of accuracy in this market due to the availability of day-ahead prices (denoted as $\tau_i^{DA}$ at timestep $i$), which are typically released 12 hours before the market opens, making it a reliable benchmark for experimental analysis. Therefore, the remainder of this paper focuses on comparing the trade-offs between the day-ahead markets and energy community use within a stacked revenue model.

Another important potential source of revenue for grid-connected batteries is though contracts to provide flexibility to grid operators, on demand, to relieve network congestion in certain locations. In the Netherlands, the Grid Operators Platform for Ancillary Services (GOPACS) \footnote{https://www.gopacs.eu/en/what-is-gopacs/} is the mechanism through which such flexibility is traded. While offering flexiblity through GOPACS is an important potential source of future revenue for grid-connected batteries, such contracts are very customized (specific to place/time and local circumstances). So far, participation in day-ahead and imbalance markets has been a much more profitable alternative for GIGA Storage and easier to model based on existing historical data, hence, in this paper, we focus on the trade-off between day-ahead market trading and community energy use. Future work could include GOPACS participation as part of the stacked revenue model of a large battery. 

\subsubsection{Fixed Energy Tariffs for Prosumers and  Energy Communities}

From the perspective of individual prosumers or energy communities, two types of tariffs have been considered: fixed and dynamic energy tariffs. Formally, we denote the buying and selling prices for consumers at timestep $i$ by $\tau^b_i$ and $\tau^s_i$, respectively. 
Fixed energy tariffs (also called flat-rate pricing), remain constant over a billing period, regardless of time or demand. Hence, under such fixed tariffs, $\tau_i^b$ and $\tau_i^s$ remain constant across all $i$. While simple and predictable, they do not encourage shifting usage from peak to off-peak hours, resulting in higher peak demand and, typically, higher overall costs.

In our study, we consider fixed tariffs ranging from €0.10 to €0.40 per kWh for imported energy. For exported energy, we consider both scenarios where no compensation is received, and scenarios where energy exports are compensated at a fixed rate of €0.10 per kWh. This choice follows closely from Eurostat \cite{EuropeanEnergyPrices} data, which shows that in 2023, Dutch energy prices were the highest in Europe, averaging €0.4523 per kWh, compared to the EU average of €0.2965. 

\subsubsection{Dynamic Energy Tariffs for Prosumers and Energy Communities}

Dynamic energy tariffs fluctuate based on demand, with higher prices during peak hours and lower prices during off-peak periods. This incentivizes consumers to use electricity when demand is low, improving grid stability. However, managing dynamic pricing can be challenging due to its inherent unpredictability. 

Dynamic tariffs follow closely from market prices, but additional costs and constraints. In our analysis, a flat tariff of €0.155/kWh is added to the day-ahead market, representing additional fees the community has to pay, such as the cost of transporting the energy. For a few time periods, the day-ahead prices can occasionally become negative (even with added taxes). In such cases, the community's buying price is capped at a minimum of zero, as negative prices are not usually part of tariffs to end-consumers and are only accessible to energy traders.  
For energy exports from the community, two scenarios are considered: one where energy exported to the grid is compensated, and one where it is not. In the compensated scenario, the selling price is set at 90\% of the buying price, capped at a maximum of €0.10/kWh. This approach aims to replicate the pricing structure offered by Frank Energy\footnote{https://www.frankenergie.nl/}, a Netherlands-based electricity supplier that provides dynamic energy tariffs.

\begin{table}[t]
\centering
\caption{Summary of parameter values used in the experiments.}
\label{tabe:param_values}
\begin{tabular}{|l l|l l|}
\hline
\textbf{Parameter} & \textbf{Value} & \textbf{Parameter} & \textbf{Value} \\
\hline
$\Delta t$           & 0.5 hours                   & $T$                     & 24 hours (48 timesteps)           \\
$\eta_c$             & 0.90                        & $\eta_d$                & 0.97                              \\
$\eta_{cd}$          & 0.87                        & $\lambda_{\text{capacity}}$ & 0.12 (€/kWh)               \\
$\lambda_{\text{charging}}$ & $10^{-7}$           & $\lambda_{\text{max cycles}}$ & 1.3                         \\
$e_{\text{max}}$     & $\infty$                    & $p_{\text{max}}$        & $0.5 \times\text{SoC}_{\text{max}}$ (kW) \\
$\text{SoC}_{\text{min}}$ & 0 kWh                  & $\text{SoC}_{\text{max}}$ & up to 7.5 MWh \\
$\text{SoC}_{\text{initial}}$ & Previous day's $\text{SoC}_{\text{EoD}}$ & $\text{SoC}_{\text{EoD}}$  & at least 0\%, 50\%, or 100\% of $\text{SoC}_{\text{max}}$ \\
$\tau^b$ & Flat: $\{0.1, 0.2, 0.3, 0.4\}$ (€/kWh) & $\tau^s$ & Flat: $\{0.0, 0.1\}$  (€/kWh) \\
        & Dynamic: $\max(0,\tau^{\text{DA}}_i + 0.155)$ &         & Dynamic: $ \min(0.9\,\tau^b_i, 0.1)$ \\
\hline
\end{tabular}
\end{table}

\subsection{Battery Control Strategies for the Energy Community}
\label{sec:meth:control_strategies_communities}

The scheduling of battery operations can significantly impact the final savings for the community. We examine two distinct control strategies to optimize battery usage, each with a different planning horizon.

The first approach is a \emph{heuristic-based greedy algorithm}, which selects the optimal action at each time step based solely on the current state, without considering future conditions. This reactive real-time control method has the advantage that it is relatively simple to implement. It does not require ``look ahead" forecasts of generation, demand, or prices, making decisions based on immediate energy consumption and generation. Nevertheless, this greedy method has been shown, initially in Norbu et al.~\cite{Norbu2021Apr}, with more detailed comparisons in~\cite{ISGT_paper_Benoit,blockchain_paper_Benoit}, to be optimal for the case of \emph{flat tariffs} (where buying/selling energy later vs. now does not make a difference), although of course, performance can degrade arbitrarily in the case of more dynamic tariffs or real-time prices. The second approach utilizes a \emph{linear optimization model} with a one-day event horizon. This method calculates the optimal battery schedule for the entire day, aiming to minimize costs or maximize savings. It requires more input data at runtime, including estimations of future energy consumption and generation for the upcoming day. The specific values used for all parameters in the simulations, as described in Sections~\ref{sec:meth:community_params}, \ref{sec:meth:energy_prices}, and~\ref{sec:meth:control_strategies_communities}, are provided in Table~\ref{tabe:param_values}.

The following subsections will provide details on each control strategy, explaining how they integrate energy consumption, generation, and price data to optimize the battery's use within the energy community.

\subsubsection{Greedy Optimization for Battery Management\label{sect:greedy}}

A straightforward, reactive control algorithm for battery management, which requires no information about future generation, demand, or prices, is a heuristic-based greedy algorithm\footnote{Note that, following standard algorithm theory~\cite{kleinberg_tardos}, the term ``greedy" here denotes that the algorithm takes the optimal decision using currently available information and resources, not modeling future states.}. The formulation proposed by Norbu et al. \cite{Norbu2021Apr} was adapted for use in this paper. This greedy algorithm prioritizes the use of the battery over the grid whenever possible. It is termed "heuristic-based" because it operates solely based on the current state without accounting for future conditions. Specifically, in scenarios of excess energy generation, the algorithm will charge the battery first, resorting to exporting energy only if the excess power exceeds the battery's specifications or if the battery lacks sufficient capacity to store all the generated energy. Conversely, during energy deficits, the algorithm will discharge the battery first, importing energy only if the required power surpasses the battery's maximum power rating or if the stored energy is insufficient to cover the deficit. This algorithm is optimal for a flat import and export tariff structure, assuming the import cost exceeds the export cost. The formal definition of the algorithm is given as Algorithm  \ref{alg:methodology-greedy-algorithm}. 


\begin{algorithm}[t]

\begin{algorithmic}[1]
\Require Generation: $g^{wind}_i$, Demand: $d_i$, Grid Price: $\tau^b_i$, $\tau^s_i$
\Require Battery Specifications: $\eta^c$, $\eta^d$, $SoC^{initial}$, $SoC^{max}$, $SoC^{min}$, $p^{max}$
\Ensure State of Charge $SoC_i$, Exported Energy $e^s_i$, Imported Energy $e^b_i$

\State Initialize $SoC_0 \gets SoC^{initial}$ \label{alg:line:greedy-1}
\State Initialize $T \gets |g^{wind}|$ \Comment{Number of timesteps}
\For{$i \gets 1, T$}
    \If{$g^{wind}_i \geq d_i$} \Comment{Excess Energy Scenario}
        \State $p^{charge}_i \gets \min\big(g^{wind}_i - d_i, p^{max}\big)$ 
        \State $SoC^{change} \gets \min\big(
        p^{charge}_i \cdot \Delta t \cdot \eta^c, 
        SoC^{max} - SoC_{i-1}\big)$ 
        \State $SoC_i \gets SoC_{i-1} + SoC^{change}$ 
        \State $e^s_i \gets (g^{wind}_i - d_i) \cdot \Delta t + SoC^{change} / \eta^c$ 
    \Else \Comment{Energy Deficit Scenario}
        \State $p^{discharge}_i \gets \min\big(d_i - g^{wind}_i, p^{max}\big)$ 
        \State $SoC^{change} \gets \min\big(
        p^{discharge}_i \cdot \Delta t / \eta^d, 
        SoC_{i-1} - SoC^{min}
        \big)$ 
        \State $SoC_i \gets SoC_{i-1} - SoC^{change} $
        \State $e^b_i \gets (d_i - g^{wind}_i) \Delta t + SoC^{change} \cdot \eta^d$ 
    \EndIf
\EndFor

\State \Comment{Post-Processing}
\State $SoC_i$: State of Charge at time $i$
\State $e^s_i$: Energy exported to grid at time $i$ with selling price of $\tau^s_i$
\State $e^b_i$: Energy imported from grid at time $i$ with buying price of $\tau^b_i$

\end{algorithmic}
\caption{Heuristic-Based Greedy Battery Control Algorithm}
\label{alg:methodology-greedy-algorithm}
\end{algorithm}

The algorithm proceeds as follows:
\begin{enumerate}
    \item \emph{Initialization}: The state of charge (SoC) is initialized to its initial value (Line 1), and the number of timesteps is determined from the generation data (Line 2).
    \item \emph{Excess Energy Scenario}: When generation exceeds demand ($g^{wind}_i \geq d_i$, Line 4), the algorithm calculates the maximum power that can be used to charge the battery without exceeding its maximum charging power (Line 5) or capacity (Line 6). The SoC is updated accordingly (Line 7). Any remaining excess energy is then exported to the grid (Line 8).
    \item \emph{Energy Deficit Scenario}: When demand exceeds generation ($g^{wind}_i < d_i$, Line 9), the algorithm calculates the maximum power that can be discharged from the battery without exceeding its maximum discharging power (Line 10) or capacity (Line 11). The SoC is updated accordingly (Line 12). Any remaining deficit is then met by importing energy from the grid (Line 13).
    \item \emph{Post-Processing}: The final state of charge, exported energy, and imported energy for each timestep are recorded for further analysis (Lines 17-19). Note that the exported energy ($e^s_i$) and imported energy ($e^b_i$) are not needed, as they can be calculated from the state of charge ($SoC_i$), generation ($g^{wind}_i$), and demand ($d_i$).

\end{enumerate}

\subsubsection{Linear Programming Optimization for Battery Management}
\label{sec:meth:linear-programing-simple}
Unlike the greedy algorithm, which makes independent decisions at each timestep, the linear programming (LP) model takes a more strategic approach by optimizing the charging schedule over a look-ahead horizon. Rather than reacting instantaneously, it considers the entire day's energy dynamics to determine a schedule that minimizes the community's energy bill.

This approach requires input data before optimization, either through forecasting or historical patterns. In this study, historical data was used for energy generation ($g^{wind}_i$), energy demand ($d_i$), and energy tariffs ($\tau_i$), as described in Sections~\ref{sec:demand_data} and ~\ref{sec:wind_generation}. How far into the future the algorithm can access data is referred to as the event horizon. A longer horizon generally leads to better scheduling, but reduces accuracy due to prediction uncertainty. For this paper, the horizon was set to one day (24 hours), aligning with current literature on battery optimization \cite{POZO2022108565} and operational practices at GIGA Storage.

\setcounter{algorithm}{0}

\begin{model}
\caption{Linear Programming Model for Battery Control - Community Simulation}\label{alg:methodology-linear-algorithm-community}
\begin{algorithmic}
    \State \textbf{Objective function:}
    \State $\text{Minimize } \sum_{i=1}^{T} \left( \tau^{b}_{i} \cdot e^{b}_{i} - \tau^{s}_{i} \cdot e^s_{i} \right) + L_1 + L_2$\Comment{(\thealgorithm.0)}

    \hfil
    
    \State \textbf{Constraints:}
    \State $SoC_0 = SoC_{initial}$ \Comment{(\thealgorithm.1)}
    \State $SoC_{i+1} = SoC_i + \eta_c \cdot p^{charge}_{i} \cdot \Delta t - \frac{p_{i}^{discharge}}{\eta_d} \cdot \Delta t$\Comment{(\thealgorithm.2)}
    \State $ p_{i}^{charge} - p_{i}^{discharge} = g^{wind}_i - d_i + e_i^b - e^s_{i}$\Comment{(\thealgorithm.3)}
    \State $\sum_{i=1}^{T} \frac{(p_{i}^{charge} \cdot \eta_c + p_{i}^{discharge} / \eta_d) \cdot \Delta t}{2 \cdot (SoC^{max} - SoC^{min})} \leq \lambda^{max\_cycles}$ \label{alg:line:linear1-4}\Comment{(\thealgorithm.4)} 
    \State $SoC_T \geq SoC^{EoD}$ \Comment{(\thealgorithm.5)}


    \hfil
    
    \State \textbf{Decision variables:}
    \State $SoC_i \in [SoC^{min},SoC^{max}] $ \Comment{State of charge at time step $i$}
    \State $p_{i}^{charge} \in [0,p^{max}] $ \Comment{Battery charging power at time step $i$}
    \State $p_{i}^{discharge}\in [0,p^{max}] $ \Comment{Battery discharging power at time step $i$} 
    \State $e^b_{i} \in [0,\infty]$ \Comment{Energy import at time step $i$}
    \State $e^s_{i}\in [0,\infty]$ \Comment{Energy export at time step $i$}

    \hfil
    \State \textbf{Regularization costs:}
    \State $L_1 = \sum_{i=1}^{T} \left( \lambda^{charging} \cdot p^{charge}_i + \lambda^{charging} \cdot p^{discharge}_i\right) $ \Comment{Penalty for using the battery}
    \State $L_2 =   \lambda^{capacity}\cdot \left( Soc^{max} - SoC_T \right)$ \Comment{Penalty for empty battery}
\end{algorithmic}

\end{model}

\hyperref[sec:meth:linear-programing-simple]{Model 1} presents the LP model used to control the battery capacity allocated to the energy community. The objective function minimizes the energy bill by balancing the cost of imported energy against revenue from exported energy (Equation 1.0). Optimization was performed using the SCIP solver, accessed via Google OR-Tools. Two regularization terms ($L_1$ and $L_2$) were included to refine model behavior, as discussed in the next section.

The model's constraints ensure the battery's state of charge (SoC) is updated correctly at each timestep. The initial SoC is predefined (Equation 1.1), and each subsequent SoC is determined based on the previous state, charging/discharging power, and respective efficiencies (Equation 1.2). The power balance constraint ensures that net energy generation minus demand aligns with energy imports, exports, and battery operations (Equation 1.3). Additionally, the maximum daily depth of charge $\lambda^{max\_cycles}$ limits excessive cycling, prolonging battery life (Equation 1.4).

The LP is particularly advantageous in scenarios where energy prices fluctuate over time. Unlike the reactive greedy method, the LP model anticipates price changes and schedules grid interactions accordingly. For example, it may purchase energy at night when prices are low and store it for later use. Additionally, the model can participate in energy trading, buying energy with the intent to sell at a higher price rather than solely for deficit coverage. This allows it to balance the community’s needs while maximizing economic benefits.

In the LP model presented above, mutually exclusive operations, such as charging and discharging, can technically occur simultaneously, which is not physically realistic. A mixed integer linear programming (MILP) formulation resolves this issue by introducing binary constraints to enforce exclusivity. However, this comes at a computational cost. In our case, dropping these binary constraints (i.e., using LP instead of MILP) does \emph{not} affect the optimality of the solution, as the objective function naturally discourages simultaneous operations (see ~\ref{appendix-comparison-milp-lp}  for a discussion of this aspect). Given this, we use the LP model for its improved computational efficiency.

\subsubsection{Regularization Techniques for Improved Scheduling\label{sect:L1_L2reg}}

LP models are impractical for large look-ahead time horizons, such as optimizing a schedule over an entire year. This is due to both the computational expense and the fact that reliable forecasts of renewable generation and load are usually available only 24 hours in advance. 
Instead, a standard approach, also employed by this study, is to run the model independently at the start of each day, generating a 24-hour ahead schedule. These schedules are then merged together by setting each day's initial battery charge to the final charge of the previous day, as defined in Equation 1.1.

While this method is computationally feasible, it introduces challenges. The primary issue is that the resulting merged schedule is not globally optimal from a whole-year perspective, even though each individual daily schedule is optimal on its own. We observed that the LP model tends to sell surplus energy rather than store it, often ending the day with an empty battery, as shown in the top-left graph of Figure~\ref{fig:methodology-SoC_all_models2}. This behavior is locally optimal since the model stores just enough surplus to cover potential shortages within the same day and exports the rest to maximize daily profits. However, this approach fails to account for energy deficits that may occur the following day (or later), when repurchasing energy is likely more expensive than the original selling price. Retaining some surplus energy would lead to a better yearly outcome. Additionally, the generated schedule tends to overuse the battery, cycling it more than necessary.

To address these issues, we introduce two regularization costs designed to refine the schedule and guide the LP model toward better long-term performance. In linear optimization, multiple solutions can yield the same objective function value, making them all technically optimal. The role of these regularization costs is to guide the search towards a preferred solution, without changing the overall objective.

The L1 cost discourages excessive battery cycling by introducing a small penalty, $\lambda^{charging}$, each time the battery is charged or discharged. This encourages solutions that minimize unnecessary battery usage. The penalty is deliberately kept small to avoid affecting the overall optimization. In our study, we set $\lambda^{charging} = 10^{-7}$. 
The L2 cost encourages the battery to retain charge at the end of the day. Due to the one-day event horizon, the LP model does not anticipate future energy needs and thus prioritizes exporting surplus energy at the end of the day. The L2 regularization introduces a small penalty, $\lambda^{capacity}$, for leaving unused battery capacity, incentivizing the model to store energy when possible. However, a potential downside is that if electricity prices fall below $\lambda^{capacity}$, the model may purchase power unnecessarily. In our study, we set $\lambda^{capacity}$ = €0.12 per kWh, selected experimentally based on performance across a large number of simulated scenarios.

The best results are achieved when both regularization techniques are applied together. This approach not only improves scheduling in all scenarios but also reduces battery wear, extending its lifespan. A detailed discussion and experimental evaluation of these regularizations can be found in Sections~\ref{sect:linear_model_comparisons} and ~\ref{sect:optimal_battery_control}.

\subsubsection{Benchmarking Against Alternative LP Approaches}
\label{sect:benchmarking_LP}

To evaluate our proposed regularization functions, we compare them against several benchmark methods. These alternative approaches were chosen because they also aim to encourage the LP model to charge the battery, but achieve this through different mechanisms rather than by modifying the objective function.

\begin{figure}[htbp]
  \centering
    \includegraphics{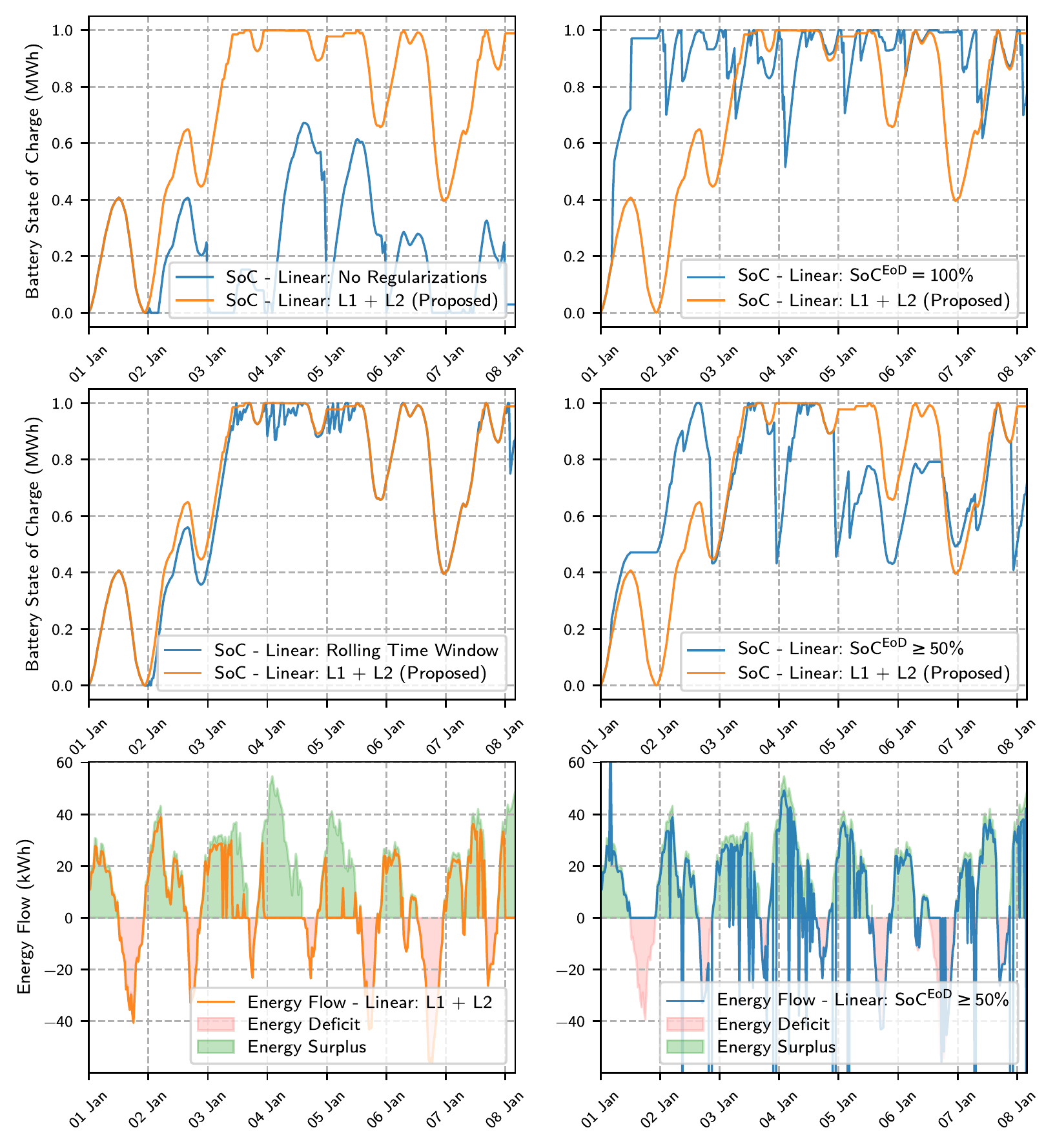}
\caption{Comparison of charging/discharging behaviour between the proposed LP with L1+L2 regularizations control model and the other benchmark linear models. The benchmark models include the unregularized LP model (top-left), the rolling-time window optimization (middle-left), the linear model with a constraint enforcing the state of charge (SoC) to be 100\% of battery capacity at the end of the day (EoD) (top-right), and the model with a constraint ensuring SoC remains at least 50\% of capacity at the EoD (middle-right). The bottom-left plot illustrates our proposed L1+L2 regularized model, highlighting energy surpluses (green), deficits (red), and battery energy flows. For comparison, the bottom-right plot presents the results for the model constrained to reach a minimum 50\% SoC at EoD, using the same visualization approach.}

  \label{fig:methodology-SoC_all_models2}
\end{figure}

The first alternative approach imposes a hard constraint on the minimum state of charge (SoC) at the end of each day (Equation 1.5). This ensures that the model concludes each day (and starts the next) with a predetermined minimum charge level. In our study, we consider two commonly used thresholds: requiring at least 50\% or 100\% SoC at the end of the day (EoD). Notably, GIGA Storage also applies the 50\% threshold in its operations. The main drawback of this method is that it forces the battery to charge even when no surplus energy is available, making it necessary to purchase energy from the grid. The SoC for both threshold levels is illustrated in Figure~\ref{fig:methodology-SoC_all_models2}, where the top-right and middle-right graphs correspond to the 100\% and 50\% SoC values, respectively. As shown in the figure, the method with a 100\% SoC at the EoD constraint keeps the battery near full capacity for the longest periods. However, this comes at a loss of efficiency, as it forces charging even during low renewable generation or high import prices.

One optimization approach, which is put forward in papers on stochastic battery control~\cite{finhold_al,malysz_al} is the rolling-time window, or rolling time horizon technique. Unlike the standard method, which applies linear optimization over a 24-hour period once per day, this technique performs optimization \emph{every hour} using a full 24-hour horizon. In practice, the model generates a complete 24-hour schedule but only implements the first hour before re-optimizing based on newly available information. This adjustment allows the model to anticipate and account for potential energy deficits beyond the end of the current day, encouraging energy storage rather than immediate selling. 

As shown in the middle-left graph in Figure \ref{fig:methodology-SoC_all_models2}, for the few days shown in this example - this method produces a schedule that closely aligns with our proposed approach with L1+L2 regularization (i.e. the charging state lines often overlap). However, a key limitation arises when energy deficits and surpluses occur far apart (beyond the length of the optimization horizon), in which case the model behaves similarly to the standard linear model without regularization.  In contrast, our proposed method with L1+L2 regularizations stores excess energy whenever unused capacity is available and the price is low, and thus is not constrained by the limited look-ahead problem. Additionally, the rolling-time window method demands significantly more computational resources, as it requires optimization to run 24 times more frequently than the standard approach in our case.

Figure~\ref{fig:methodology-SoC_all_models2} is an example that provides a view of how these different methods for controlling the battery compare to the L1+L2 regularization method, for the horizon of a few days. The full results comparing the methods for the entire dataset, are discussed in Section~\ref{sect:optimal_battery_control} and Figure~\ref{fig:appendix:comparison_flat_tariff}.

\subsection{\label{sec:market-sim}Battery Control Strategies for the Energy Market}

The spot market simulation will also employ a linear model as described in \hyperref[alg:methodology-linear-market]{Model 2} (implemented as an open source package in \cite{Jip_battery-trading-benchmark}). This model operates on a daily time horizon, similar to its community counterpart, and is executed independently for each day of the year. However, \hyperref[alg:methodology-linear-market]{Model 2} is simpler than \hyperref[sec:meth:linear-programing-simple]{Model 1}, as it does not account for demand and generation. The battery is always charged using imported energy, and discharged energy is always exported. Another notable difference is in the objective function, in this case, the profits are maximized. Formulating the problem as profit maximization or cost minimization results in equivalent LP formulations, the only difference being the sign of the objective function.

The only input for the model is the electricity prices at each time step. The historical prices from the day-ahead market ($\tau^{DA}_i$) were used for both the buying and selling prices in the market simulation. The model output will be the optimal schedule, i.e., the schedule that maximizes profits. The constraints ensure the proper functioning of the model: the state of charge (SoC) starts at an initial value and updates based on the charging and discharging powers at each timestep. The imported energy equals the charging power times the length of the time step. An identical constraint is also added for the exported energy. Additionally, the model ensures that the depth of charge does not exceed a maximum daily cycle limit. The decision variables include the state of charge at each time step, the battery charging and discharging power, and the energy imported and exported at each time step.

\begin{model}
    
\caption{Linear Programming Model for Battery Control - Market Simulation}
\label{alg:methodology-linear-market}
\begin{algorithmic}

    \State \textbf{Objective function:}
    \State $\text{Maximize } \sum_{i=1}^{T} \left( \tau^{DA}_{i} \cdot e^s_{i} - \tau^{DA}_{i} \cdot e^{b}_{i}  \right)$\Comment{(2.0)}
    
    \State \textbf{Constraints:}
    \State $SoC_0 = SoC_{initial}$\Comment{(2.1)}
    \State $SoC_{i+1} = SoC_i + \eta_{cd} \cdot p^{charge}_{i} \cdot \Delta t - p_{i}^{discharge} \cdot \Delta t$\Comment{(2.2)}
    \State $e^b_{i} = p^{charge}_{i} \cdot \Delta t $\Comment{(2.3)}
    \State $e^s_{i} = p_{i}^{discharge} \cdot \Delta t$\Comment{(2.4)}

    \State $\sum_{i=1}^{T} \frac{(p_{i}^{charge} \cdot \eta_{cd} + p_{i}^{discharge}) \cdot \Delta t}{2 \cdot (SoC^{max} - SoC^{min})} \leq \lambda^{max\_cycles}$ \label{alg:line:linear1-4:bis} \Comment{(\thealgorithm.5)}
    \State $SoC_T \geq SoC^{EoD}$ \Comment{(\thealgorithm.6)}

    \State \textbf{Decision variables:}
    \State $SoC_i \in [SoC^{min},SoC^{max}] $ \Comment{State of charge at time step $i$}
    \State $p_{i}^{charge} \in [0,p^{max}] $ \Comment{Battery charging power at time step $i$}
    \State $p_{i}^{discharge}\in [0,p^{max}] $ \Comment{Battery discharging power at time step $i$}
    \State $e^b_{i} \in [0,\infty]$ \Comment{Energy import at time step $i$}
    \State $e^s_{i}\in [0,\infty]$ \Comment{Energy export at time step $i$}

\end{algorithmic}
\end{model}

\subsection{Estimating the Rental Price Range for Different Battery Capacities}

To determine a financially viable rental price for battery storage, we analyze the problem from two perspectives: the energy community, which determines the maximum price they are willing to pay, and the battery operator, which sets the minimum price they are willing to accept. The full process is illustrated in Figure~\ref{fig:graphical_methodology}. To establish these price bounds, we conduct a year-long simulation in which we iteratively evaluate multiple battery rental capacities. 

\subsubsection{Maximum Battery Rental Price (Community Perspective)}
To determine the maximum rental price the community is willing to pay, we first analyze the community’s yearly electricity costs under different battery capacities. This analysis excludes the battery cost itself and considers only the electricity bill and the amortized cost of the wind turbine. By comparing the costs with and without battery storage, we quantify the financial savings for the community, which represents the upper limit of an economically viable rental price.

The analysis is conducted across multiple electricity tariff scenarios to account for different market conditions. Specifically, in the flat tariff case, we simulate four different buying prices (€0.4, €0.3, €0.2, €0.1 per kWh) and two selling prices (€0.0, €0.1 per kWh), resulting in a set of flat tariff scenarios. This enables us to evaluate how varying energy prices influence community savings and determine a realistic benchmark for the maximum battery rental price.

\subsubsection{Minimum Battery Rental Price (Battery Operator Perspective)}

To determine the minimum rental price the battery operator is willing to accept, we analyze the potential earnings from trading the same battery capacity on the day-ahead energy market. Since we consider a stacked-revenue model, the battery is partially allocated to market trading, while the remaining capacity can be rented out to the community. This allocation strategy ensures that the battery is used in the most economically efficient way.

In practice, the majority of the battery capacity is dedicated to the day-ahead market, as this is generally the most profitable use. However, as more capacity is allocated to market trading, its marginal profitability decreases. This diminishing return presents an opportunity to allocate the least profitable portion of the battery to alternative uses, such as renting part of the capacity to an energy community.

The analysis is based on a 7.5 MWh battery operated by GIGA Storage. First, we simulate the total annual revenue the entire system could generate by participating exclusively in the day-ahead market. Next, we iteratively allocate increasing portions of the battery to rental use and calculate the corresponding loss in market profits. This lost revenue, known as the opportunity cost in economic terminology, sets the lower bound for a financially viable rental price. To ensure realistic estimates, we use historical electricity prices from the Netherlands (2023) and apply a linear profitability model.

\newcommand{\pricefz}{$\tau^b = 0.4$ and $\tau^s = 0.0$}
\newcommand{\pricefo}{$\tau^b = 0.4$ and $\tau^s = 0.1$}
\newcommand{\pricetz}{$\tau^b = 0.3$ and $\tau^s = 0.0$}
\newcommand{\priceto}{$\tau^b = 0.3$ and $\tau^s = 0.1$}
\newcommand{\pricedz}{$\tau^b = 0.2$ and $\tau^s = 0.0$}
\newcommand{\pricedo}{$\tau^b = 0.2$ and $\tau^s = 0.1$}
\newcommand{\priceoz}{$\tau^b = 0.1$ and $\tau^s = 0.0$}
\newcommand{\priceoo}{$\tau^b = 0.1$ and $\tau^s = 0.1$}

\section{\label{cha:exp}Experimental Results}

This section presents the experimental results, verifying our approach using real-world data from the Netherlands and the UK. We begin by analyzing feasible rental price ranges for different battery capacities. Next, we determine the optimal battery rental capacity and wind turbine installation size to maximize cost savings. We then perform an ablation study on our proposed method to evaluate the impact of its individual constraints in the LP optimization. Finally, we benchmark our method against alternative approaches to assess its overall performance.

\subsection{Determining the Profitability Cut-Off Points for Battery Rental Prices \label{sec:methodology:cut-off-points}}

\begin{figure}[t]
    \centering
    \centerline{
    \includegraphics[]{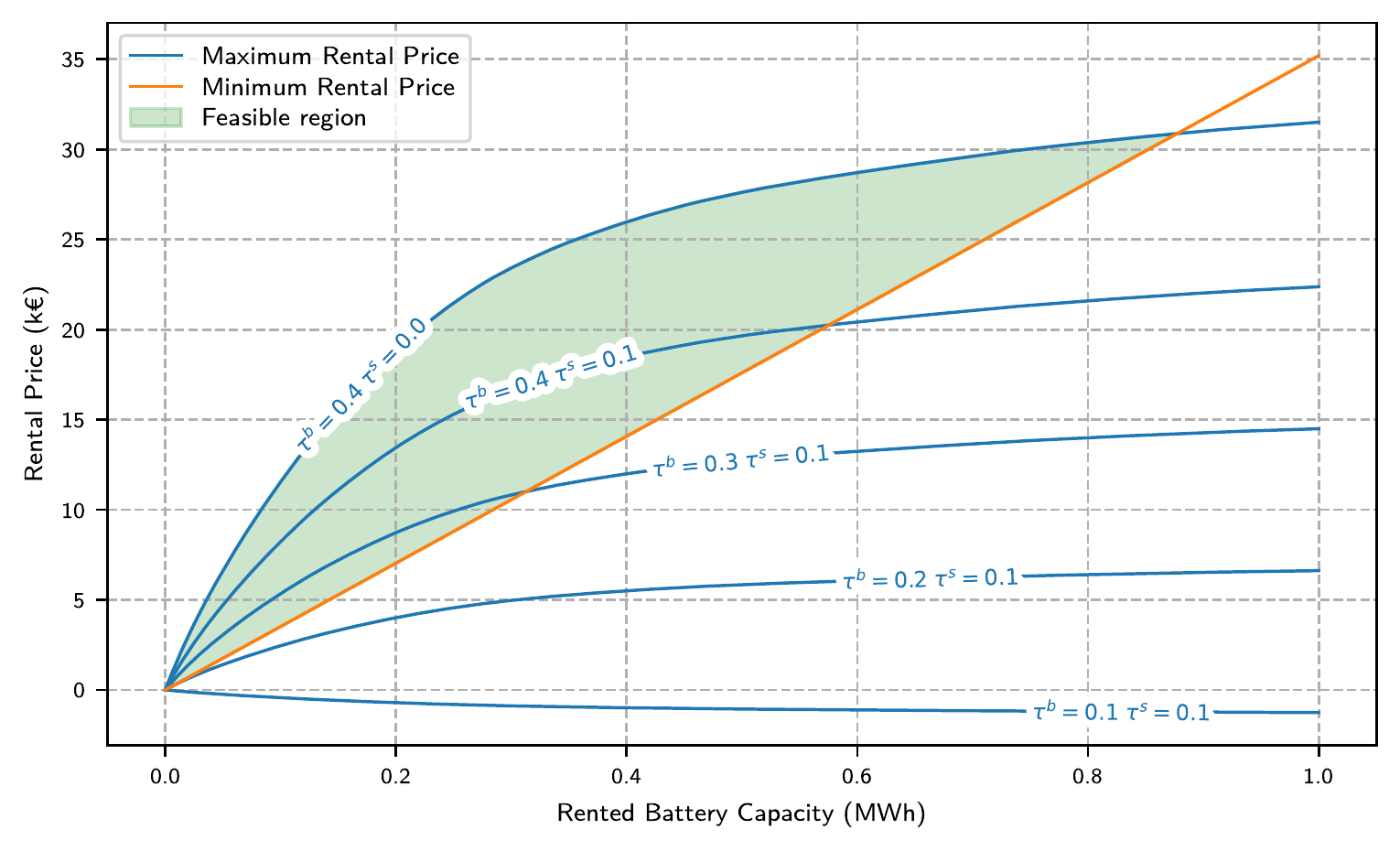}
    }
    \caption{Minimum and maximum feasible rental prices for different battery capacities under various flat tariff scenarios (in € per kWh). Rental remains profitable for both the energy community and the battery operator in the green area, where the maximum price accepted by the community exceeds the minimum price required by the battery operator.}

    \label{fig:experiments-min-max-battery-price}
\end{figure}

In this section, we analyze which battery capacities would be feasible to rent and at what price. Specifically, both the maximum and minimum prices (from the energy community and battery operator, respectively) are shown in Figure \ref{fig:experiments-min-max-battery-price}. Multiple maximum prices exist depending on the flat-tariff scenario used. The feasible area, i.e., the area where the maximum price is greater than the minimum price is depicted in green. Selecting any combination of battery size and yearly rental price from the green area will result in both a lower bill for the community and an increase in profit for the company. The results show that renting out the battery can be profitable for both parties up to a capacity of 900 kWh if a flat import tariff of forty cents $\tau^b = 0.4$ (which was common in the Netherlands in 2023) is used and no export is allowed ($\tau^s = 0$). As the graph shows, the methodology would also be applicable, however, in lower per kWh tariff scenarios, encountered in other EU countries.

\subsection{Optimal Co-Sizing of Battery and Wind Generation Capacity}

In this subsection we will examine the co-optimization problem for the battery sizing and cost, in conjunction with the optimal size for wind generation in the community, such that the joint system yields maximum savings for the community. Throughout this section, the cost of the battery is included in the community's bill, with the rental price set at the minimum level that is still profitable for the battery operator.

\subsubsection{Optimal Battery Capacity for Maximizing Savings \label{sec:methodology:optimal_battery_capacity}}
\begin{figure}
\begin{subfigure}{.5\textwidth}
    \centering
    \caption{The annual costs}
    \includegraphics[]{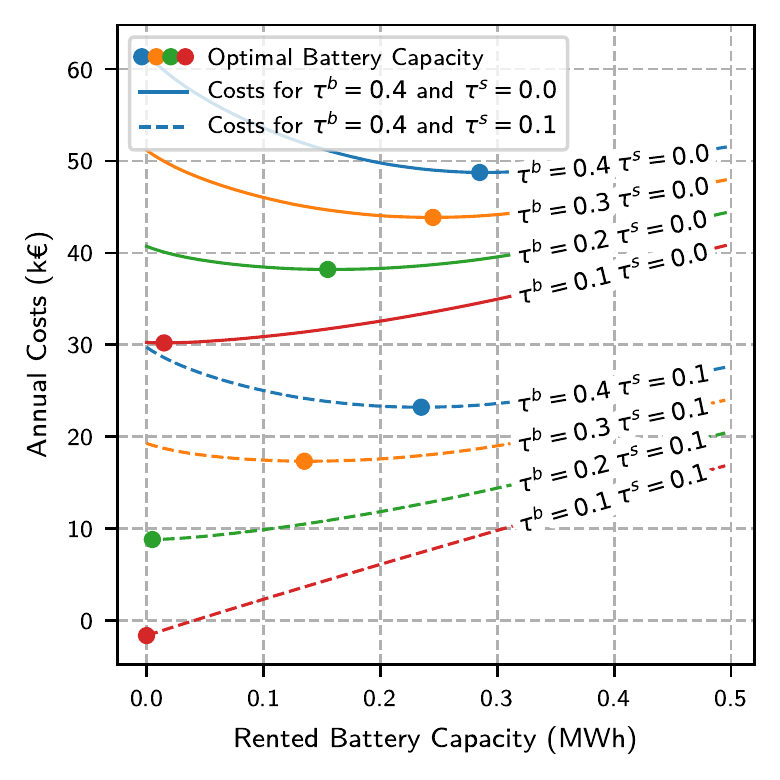}
    \label{fig:experiments-optimal-battery-size}
\end{subfigure}%
\begin{subfigure}{.5\textwidth}
    \centering
    \caption{The annual savings}
     \includegraphics[]{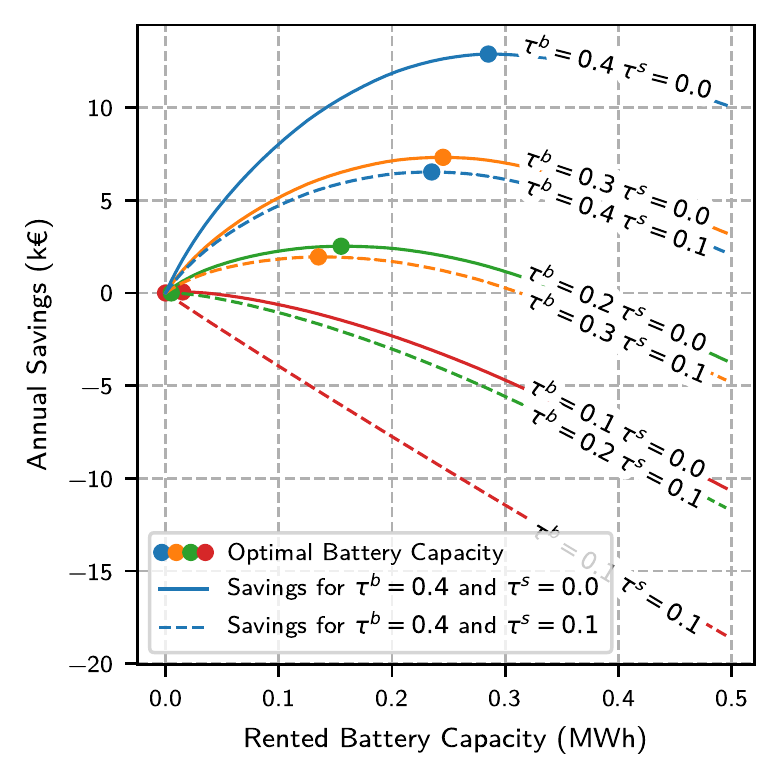}
    \label{fig:experiments-optimal-battery-size_savings}
\end{subfigure}
\caption{Annual costs and annual savings for the community, including the battery rental price, across different flat tariff scenarios (in € per kWh). The optimal battery capacity for each scenario is indicated, representing the size that minimizes costs or maximizes savings for the community.}
\end{figure}

To determine the battery capacity that maximizes benefits for both the community and the battery owner, the optimal battery size was calculated for each scenario and is shown in Figure \ref{fig:experiments-optimal-battery-size}. The optimal points for each scenario are indicated by dots on the coloured lines. These points represent the optimal battery capacity that corresponds to the lowest achievable bill for the community, typically at the inflection points of the lines. We can see from the results that for most scenarios, a capacity between 100 to 300 kWh is optimal. For three of the studied scenarios, the optimum capacity is zero, meaning it is not profitable to rent any battery capacity in those scenarios, and the community can satisfy their needs by only using the grid. Note, however, all these scenarios correspond to a buying price of either €0.10 or €0.20/kWh -- whereas in the Netherlands, an average retail tariff of €0.40/kWh was frequent in 2023. This suggests that the proposed methodology would be advantageous in the Netherlands and many EU countries, where retail tariffs are high.

To further investigate this problem, we compute the financial savings, which can provide a clearer view. The savings can be calculated directly from the cost and can be seen in Figure \ref{fig:experiments-optimal-battery-size_savings} for all studied scenarios. Note that the optimal values are the same regardless of the metric used, the only difference is that we either minimize or maximize, depending on the case. We can see that adding 10 cents to the export price has roughly the same effect as subtracting 10 cents from the buying price. This explains why the optimal values between the corresponding cases are almost identical.

Table \ref{tab:optimal_battery_size} shows the total cost, total savings, and battery capacity for the optimal case for each of the scenarios. The results show that in most cases, renting the battery is advantageous, with savings ranging from €2,516 to €12,874 for a 200-household community equipped with a 330 kW wind turbine, depending on the type of flat tariff used. The high range of savings indicates that the feasibility of the proposed methodology is highly dependent on the import tariff used. The export tariff is not as important in this case, as it can be absorbed into the import price. 

\begin{table}[t]
\centering
\caption{The minimum costs, maximum savings, and optimal battery capacity for each of the analyzed flat tariff scenarios.}
\resizebox{1\textwidth}{!}{%
\begin{tabular}{|l||rr||rr||rr||rr||}
\hline
Import Tariff (in € per kWh)&
  \multicolumn{2}{c||}{$\tau^b = 0.4$} &
  \multicolumn{2}{c||}{$\tau^b = 0.3$} &
  \multicolumn{2}{c||}{$\tau^b = 0.2$} &
  \multicolumn{2}{c||}{$\tau^b = 0.1$} \\ \hline
Export Tariff (in € per kWh)&
  \multicolumn{1}{r|}{$\tau^s = 0.0$} &
  $\tau^s = 0.1$ &
  \multicolumn{1}{r|}{$\tau^s = 0.0$} &
  $\tau^s = 0.1$ &
  \multicolumn{1}{r|}{$\tau^s = 0.0$} &
  $\tau^s = 0.1$ &
  \multicolumn{1}{r|}{$\tau^s = 0.0$} &
  $\tau^s = 0.1$ \\ \hline
Annual Costs (in €) &
  \multicolumn{1}{r|}{48762} &
  23214&
  \multicolumn{1}{r|}{43868} &
  17357&
  \multicolumn{1}{r|}{38202} &
  8814&
  \multicolumn{1}{r|}{30259} &
  -1645\\ \hline
Annual Savings (in €) &
  \multicolumn{1}{r|}{12874} &
  6518&
  \multicolumn{1}{r|}{7310} &
  1916 &
  \multicolumn{1}{r|}{2516} &
  0&
  \multicolumn{1}{r|}{0} &
  0\\ \hline
Battery Capacity (in kWh) &
  \multicolumn{1}{r|}{280} &
  240&
  \multicolumn{1}{r|}{240} &
  120&
  \multicolumn{1}{r|}{160} &
  0 &
  \multicolumn{1}{r|}{0} &
  0\\ \hline
\end{tabular}%
}
\label{tab:optimal_battery_size}
\end{table}

\subsubsection{Optimal Wind Turbine Capacity for Maximizing Savings}
\label{sect:renewable_sizing_optimization}

\begin{figure}
\begin{subfigure}{.5\textwidth}
    \centering
    \caption{Optimal annual costs}
\includegraphics[]{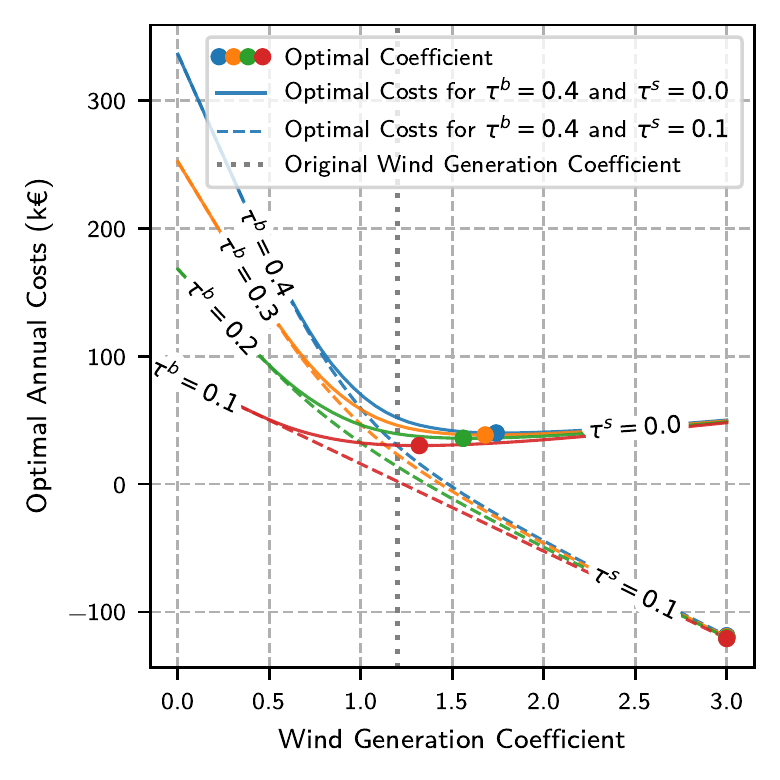}
    \label{fig:experiments-optimal-scale-costs}
\end{subfigure}%
\begin{subfigure}{.5\textwidth}
    \centering
    \caption{Optimal battery capacity}
    \includegraphics[]{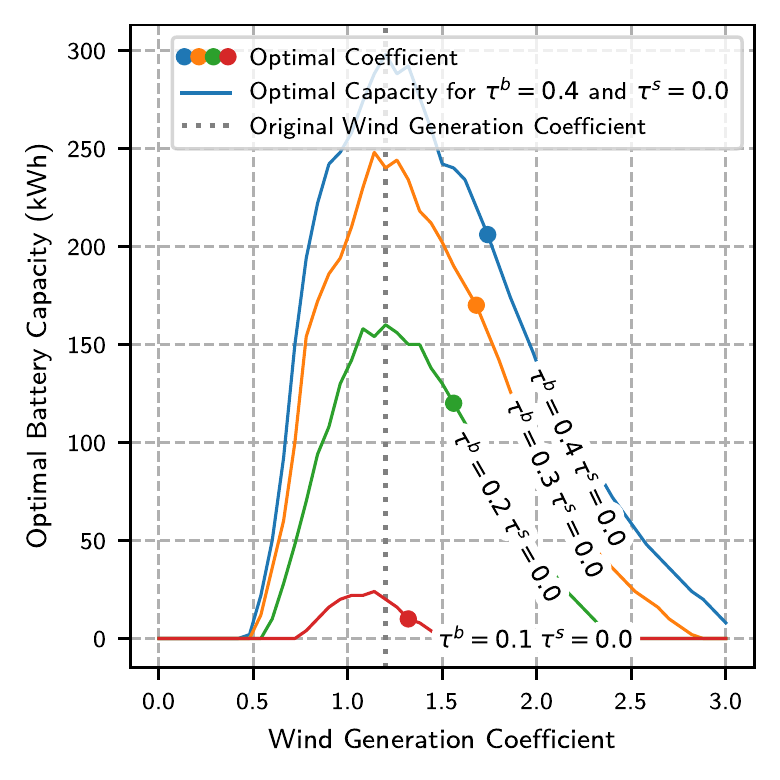}
    \label{fig:experiments-optimal-scale-capacities-without-export}
\end{subfigure}
\caption{Optimal annual costs and optimal battery capacity for different wind generation coefficients under various flat tariff scenarios (in € per kWh). The optimal generation coefficient for each scenario is indicated, representing the value that minimizes costs.}
\label{fig:experiments:optimale-scale}
\end{figure}

To determine the optimal wind turbine capacity that maximizes savings for our community of 200 households, we extend our methodology by incorporating the renewable generation coefficient, as introduced in Section \ref{sec:wind_generation}. This coefficient represents the fraction of the community’s annual electricity demand that is generated by the wind turbine, regardless of when consumption occurs (e.g., a coefficient of 0.5 indicates that the turbine produces energy equivalent to half of the community’s yearly demand). Since generation and consumption may not align, battery storage is necessary to optimize usage. By applying the full methodology, we evaluate the annual cost and determine the optimal battery capacity across different generation coefficients.

The minimum costs for each coefficient are shown in Figure \ref{fig:experiments-optimal-scale-costs}, where colors represent different buying prices, and solid and dashed lines indicate cases without and with export tariffs, respectively. In the scenario without export ($\tau^s$ = €0/kWh), the optimal coefficient ranges from 1.3 to 1.7, depending on the flat import tariff, which is close to the original Enercon wind turbine coefficient of 1.2. Notably, with a flat import tariff of €0.40/kWh, the community’s energy costs drop from €300,000 to approximately €30,000 solely by installing the original wind turbine, though further savings could be achieved with a larger turbine. In the scenario allowing paid export to the grid ($\tau^s$ = €0.1/kWh), represented by the dashed lines, increasing the turbine size consistently leads to higher profit. However, this assumes unlimited export capacity, which is not always a realistic assumption, as curtailment would limit exports beyond a certain point in practice.

We also examine the optimal battery capacity for each level of wind generation coefficient. Results for the flat tariff scheme without export are shown in Figure \ref{fig:experiments-optimal-scale-capacities-without-export} (results for the export case are similar and have been omitted). The findings reveal that optimal battery capacity initially increases with turbine size until reaching an inflection point, after which it begins to decline. Before this point, the community is under-generating and relies on battery storage to shift excess generation to later times. Beyond the inflection point, generation more frequently coincides with demand, reducing the need for storage. Interestingly, for the power curve and scaling used in this analysis (as presented in  Appendix~\ref{appendix:wind_generation}), the inflection point closely aligns with the optimal coefficient. This indicates that the largest battery capacity is required for a wind turbine size around 1.2, rather than at 1.0 (where annual generation matches annual consumption), and is thus the renewable generation setting used in all other battery experiments in this paper.


\begin{figure}[t]
\begin{subfigure}{.5\textwidth}
    \centering
    \caption{Cumulative battery degradation}
    \includegraphics[]{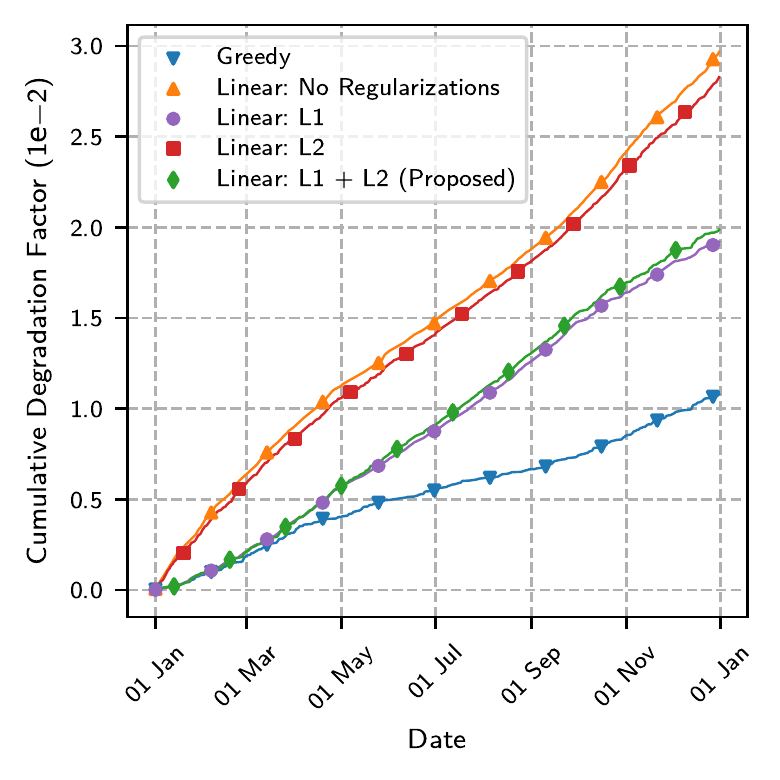}
    \label{fig:degradation_dynamic}
\end{subfigure}%
\begin{subfigure}{.5\textwidth}
    \centering
    \caption{Cumulative community savings}
    \includegraphics[]{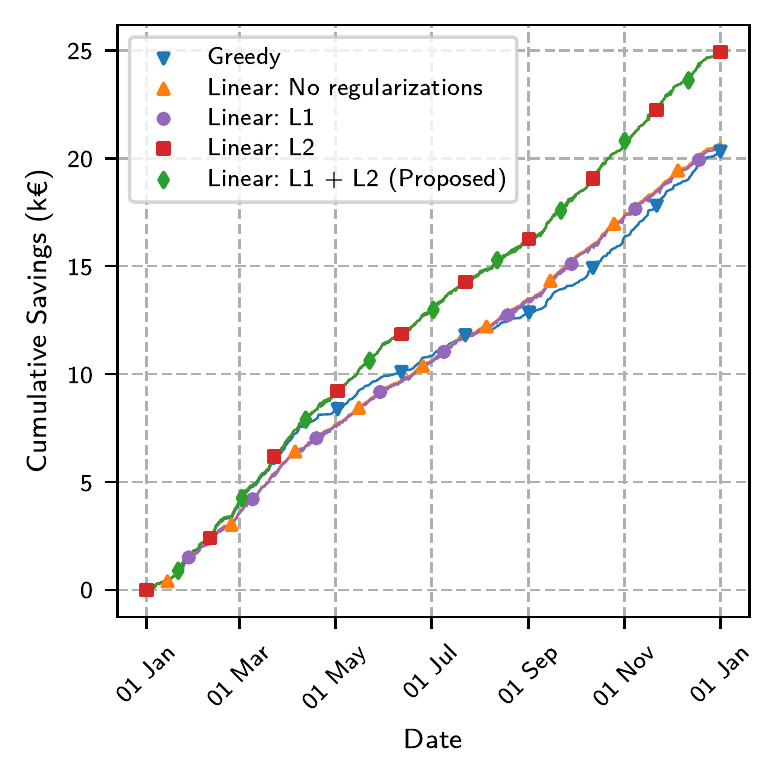}
    \label{fig:cost_savings_dynamic}
\end{subfigure}
\caption{Comparison of cumulative battery degradation and cumulative savings over the year 2023 for different battery control models under dynamic tariffs. The results demonstrate how L1 and L2 regularization influence battery wear and savings, compared to the heuristic-based greedy algorithm and the unregularized linear model.}

\label{fig:degradation_and_cost_savings_dynamic}
\end{figure}

\subsection{Impact of Our Regularization Methods on Battery Degradation and Cost Savings}
\label{sect:linear_model_comparisons}

In this section, we conduct an ablation study to evaluate the effect of the two proposed regularization techniques, L1 and L2 on the LP optimization. Specifically, we analyze the individual and combined effects of our two regularization techniques, comparing them against the baseline models: the heuristic-based greedy algorithm and the unregularized linear model. Our evaluation focuses on how these models influence battery scheduling by assessing two key performance metrics: \textbf{community cost savings} (relative to a scenario with no battery storage) and \textbf{cumulative battery degradation}. Here, cumulative degradation refers to the fraction of the remaining useful lifetime (RUL) consumed by the battery, quantified using the well-established \textit{rainflow counting} algorithm~\cite{rainflow_counting_original,Norbu2021Apr,norbu_constraints}. For completeness, a formal description of this algorithm is provided in Appendix~\ref{app_subsec:battery_degradation}.



To aid intuition, we present the cumulative growth of both key performance indicators: cost savings and battery degradation, over the course of a year (2023). All experiments assume a fixed battery capacity of 1 MWh and evaluate the effects of L1 and L2 regularization under a dynamic tariff scheme, with zero export payments (i.e. no compensation for energy exported to the grid).


Figure~\ref{fig:degradation_and_cost_savings_dynamic} presents the results, plotted cumulatively across the days in 2023. 
Note that the heuristic-based greedy algorithm results in the lowest battery degradation, while the unregularized linear model leads to the highest. The reason for this is that the greedy model is the most conservative in terms of charge/discharge cycles for the battery, although, of course, such a conservative strategy may not be the optimal one in terms of cost savings. As shown in (Figure~\ref{fig:cost_savings_dynamic}), the linear model with either L1 or with both L1+L2 achieves the best long-term savings. In summary, adding L1 regularization reduces battery degradation without affecting cost savings, whereas L2 regularization increases savings but does not further accelerate degradation. When both L1 and L2 regularization are applied together, savings reach the same level as with L2 alone, while degradation remains lower than with L2 alone. 
Thus, the best performance is achieved when both L1 and L2 regularization are applied together, as this combination balances degradation reduction with maximum savings.

\subsection{Optimal Battery Control}
\label{sect:optimal_battery_control}

In this section, we shift focus to the community's yearly costs across various battery capacity allocations, as explored in Sections ~\ref{sec:methodology:cut-off-points} and \ref{sec:methodology:optimal_battery_capacity}. We evaluate the performance of greedy and linear models under two tariff scenarios: flat tariffs and dynamic tariffs, both with an export tariff ($\tau^s = 0.1$).

We compare 6 options for the control of the battery in this setting:
\begin{itemize}[itemsep=0em]
\item The ``greedy" control heuristic (c.f. Sect.~\ref{sect:greedy})
\item The ``simple" day-ahead linear model, i.e.  without any regularizations (c.f. Sect.\ref{sec:meth:linear-programing-simple})
\item The proposed linear optimization model, which incorporates both L1 + L2 regularizations (Sect.~\ref{sect:L1_L2reg})
\end{itemize}
Note that all linear models use a 24-hour planning horizon. In addition, we consider the three extensions to linear models introduced in Section~\ref{sect:benchmarking_LP}:
\begin{itemize}[itemsep=0em]
\item The LP model with a constraint requiring the battery to be fully charged at the end of the day (EoD) 
\item The LP model including a constraint that the battery maintains at least a 50\% charge at the EoD 
\item The rolling-window model that re-optimizes every hour, whereas all other linear models operate with a single optimization at the start of each day. 
\end{itemize}

\subsubsection{Flat Tariffs}

First, we analyze the scenario involving flat tariffs. Figure \ref{fig:appendix:comparison_flat_with_export} presents the results for a constant import tariff of forty cents ($\tau^b = 0.4$ \eurpkwh) and a constant export price of ten cents ($\tau^b = 0.1$ \eurpkwh). In this case, our proposed method achieved the lowest yearly cost, matching the performance of the greedy model, which is optimal under these conditions. All other linear methods performed considerably worse.

These results align with expectations. The greedy battery control algorithm proposed by Norbu et al.\cite{Norbu2021Apr} is known to be optimal for flat tariffs when the buying price exceeds the selling price ($\tau^b > \tau^s$). This is because, under flat import and export tariffs, the decision to sell or buy a quantity of energy now versus at a future time is indifferent. Therefore, an optimal strategy is to make greedy decisions at each time step based on the currently available capacity. While this heuristic is not suitable for dynamic or market-based tariffs, where prices vary over time, our proposed method successfully replicates its behavior by incorporating L2 regularization into the linear model. As a result, we achieve optimal cost performance while maintaining the advantages of a linear formulation.

\subsubsection{Dynamic Tariffs}

\begin{figure}
\begin{subfigure}{.5\textwidth}
    \centering
    \caption{Annual costs under a flat tariff}
    \includegraphics[]{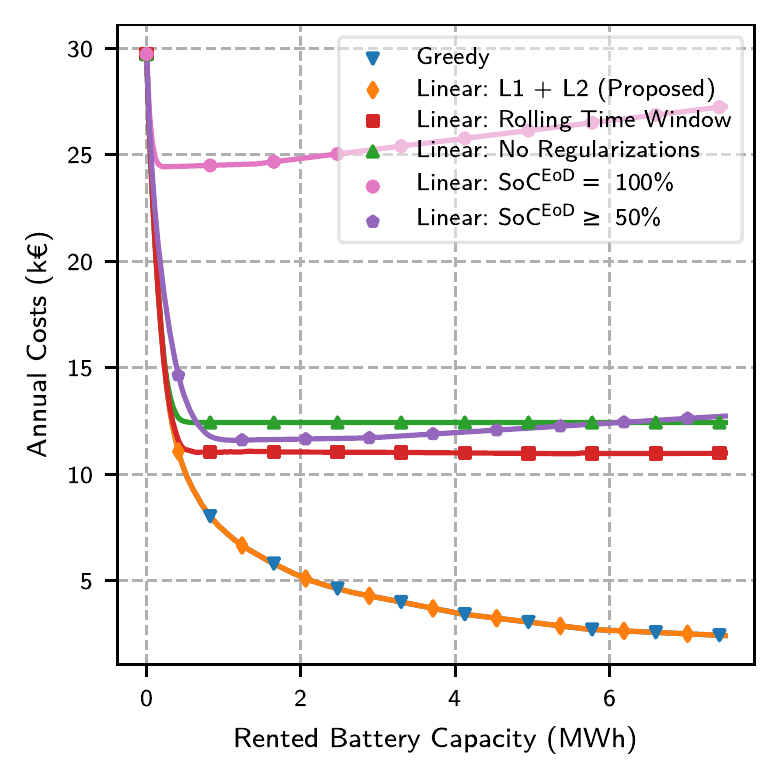}
    \label{fig:appendix:comparison_flat_with_export}
\end{subfigure}%
\begin{subfigure}{.5\textwidth}
    \centering
    \caption{Annual costs under a dynamic tariff}
    \includegraphics[]{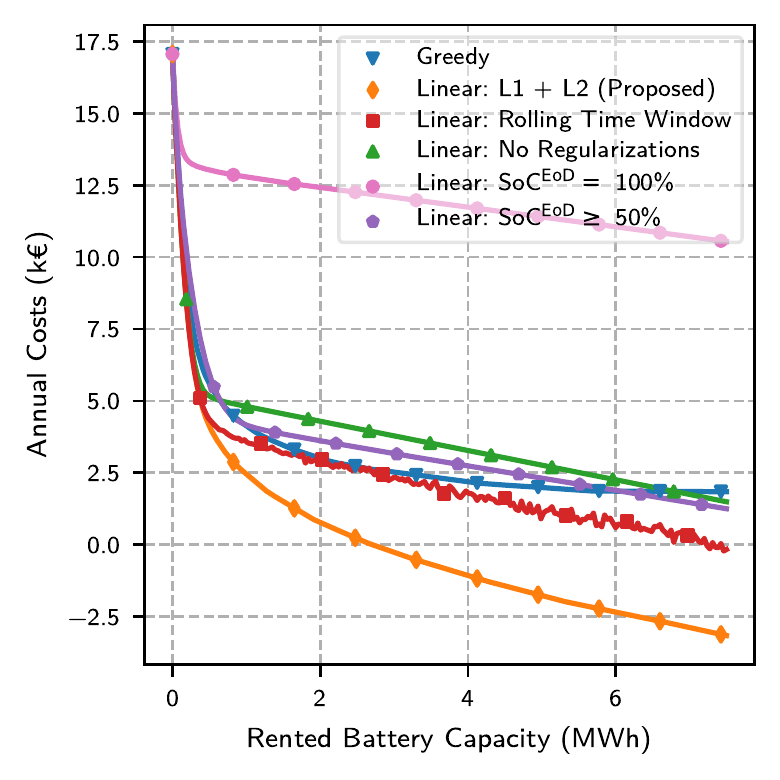}
    \label{fig:appendix:comparison_dynamic_with_export}
\end{subfigure}
\caption{Comparison of annual costs across different battery control models under flat and dynamic tariffs. The results highlight the effectiveness of our proposed model, which incorporates L1 and L2 regularization, in minimizing costs across both tariff structures.}
\label{fig:appendix:comparison_flat_tariff}
\end{figure}

Next, we evaluate the performance of the models under dynamic tariffs, which is, arguably, the most realistic case. Figure~\ref{fig:appendix:comparison_dynamic_with_export} presents the yearly costs for varying battery capacities in a dynamic tariff scenario with export. The results clearly demonstrate that our proposed method, which incorporates both L1 and L2 regularization, outperforms all other studied approaches.

As expected, the greedy strategy performs worse than both our proposed method and the rolling time window method, confirming its unsuitability for dynamic tariffs. However, it still performs very close to three of the other models, including the linear model without regularization. This suggests that while the greedy strategy does not take advantage of fluctuating prices, prioritizing the reuse of locally produced energy can be a viable approach, even under dynamic tariffs.
In fact, the proposed L1+L2 regularization can be seen as having a ``greedy" component, through its preference to save battery capacity, if current export prices are too low. 


In terms of comparison to the other strategies, the proposed linear method with L1+L2 regularization clearly outperforms them, achieving the lowest costs.  
The two linear models that enforce a hard constraint to maintain the SoC at the end of the day above a certain threshold perform relatively poorly. As expected, the method that forces a recharge of the battery to 100\% capacity each day does the worst, while the method sometimes used in practice as a heuristic (enforcing a charging of the battery to $SoC \geq 50\%$ at the end of the day) performs similarly to LP with no regularization or the greedy heuristic, for most battery capacities.

As expected, the method that does the best in the benchmarks is the method with the rolling time window (or rolling horizon) heuristic. Its performance is close to the method we propose with L1+L2 regularization. Still, there are cases when it underperforms, when there are events with a significant energy deficit just beyond its look-ahead horizon. Recall also that this method performs 24 times more computations than the other methods, hence, it also has a higher computation cost.

\section{\label{cha:disc}Discussion}

The integration of renewable energy sources and the increasing demand for energy storage systems have driven the development of innovative methodologies to optimize the use of Battery Energy Storage Systems (BESS). This discussion explores the methodology proposed for battery owners to offer their assets as a service to energy communities, thereby enhancing their revenue streams.

In this section, we will delve into the implications and findings of our research on implementing Battery as a Service (BaaS) for energy communities. We will first analyze the feasibility and profitability of the proposed model. Then we will examine how different optimization methods perform under various tariff types and market conditions. By exploring the results from different perspectives, we aim to provide a comprehensive understanding of the potential benefits and challenges associated with the proposed approach.

\subsection{Feasibility of Battery as a Service for Energy Communities}

The main contribution of our work is the proposed methodology, which enables battery owners to include energy communities as an additional revenue stream in their existing stacked revenue models. The methodology is based on the fact that most Battery Energy Storage Systems (BESS) generate a profit by participating in energy markets, such as the day-ahead market. By adding energy communities as a revenue stream, battery owners can rent out their storage capacity to communities, thereby diversifying their income sources. This approach not only helps the community lower their energy bills but also reduces their reliance on the grid.

An analysis of the proposed methodology using real data from a case study from the Netherlands for the year 2023, has shown that, with a flat import tariff of forty cents, a renewable energy community of 200 households equipped with a 330KW wind turbine behind a neighborhood transformer could generate an annual profit between €6,518 and €12,874, depending on whether exporting to the grid was allowed or not. This reflects the financial savings of the community after subtracting the battery rental cost which, in this case, is set to the potential lost profit from using that capacity for day-ahead trading. 

Hence, our methodology demonstrates that it is possible to replace some of the day-ahead market profits with a flat, guaranteed rental fee for the community. A crucial aspect of this approach is setting the rental price for battery capacity. We have outlined the region of feasible prices and battery capacities, as shown in Figure \ref{fig:experiments-min-max-battery-price}. While this paper does not aim to determine the exact price level that should be selected, this could be negotiated directly between the parties or determined through an automated, agent-mediated negotiation process, as proposed in ~\cite{etukudor2020automated}.
In practice, any price level in the feasible range would result in benefits for both parties. At the maximum price, the community would not save anything but might still be incentivized by the prospect of increasing their self-reliance. At the minimum price, the company would earn exactly as much as it would from the linear models on the energy markets. However, as mentioned previously, this optimal profit is hard to obtain in practice (because forecasts of future prices on the energy market are uncertain/more volatile), which means that even using the minimum rental price would likely result in financial gains for the battery operator company.

We also demonstrated how to determine the optimal battery size. For a typical energy community of around 200 households in the Netherlands, the ideal battery capacity was found to be between 240 kWh and 280 kWh, depending on whether or not energy export was allowed. As expected, higher energy prices and restrictions on exports increased the required battery size. Interestingly, even with a relatively low electricity tariff of €0.2 per kWh, the battery still reduced the community’s overall energy costs. This suggests that the model could also be viable in other European countries with lower electricity prices than the Netherlands.

\subsection{Performance of Battery Control Models Across Different Tariff Types}

In terms of model performance, we have analyzed how different control strategies perform under two distinct retail tariff types: flat and dynamic. Our method, the linear model using the L1 and L2 regularization costs, achieved the best performance in both scenarios. More specific, in the case of the flat tariff, our method matched the performance of the greedy model, which is known to be optimal if prices remain constant over time.  In the case of the dynamic tariff, our method greatly outperforms all other studied control algorithms.

Furthermore, our analysis highlights that, in an energy community setting, the unregularized linear model with a one-day time horizon is a poor choice for battery control. On one hand, we observed that this model frequently generates schedules that involve unnecessary charging and discharging cycles, without yielding any clear financial benefits. While these schedules are optimal from the optimizer's perspective, they contribute to excessive battery degradation. This occurs because multiple schedules can achieve the same optimal cost, some of which cause more wear on the battery than others. To address this, our proposed L1 regularization introduces a small penalty for battery usage, guiding the optimizer toward an optimal solution that also minimizes battery wear.

On the other hand, the resulting schedules are also suboptimal in terms of annual profitability, performing worse than the much simpler heuristic greedy real-time control algorithm, which does not account for electricity prices (as shown in Sections 4.3 and 4.4). We observed that the linear model, when executed daily with a 24-hour time horizon, tends to sell surplus energy at the end of the day rather than storing it. This limitation arises from the model’s restricted time horizon, as it lacks access to the next day's load and therefore cannot anticipate the long-term benefits of storing excess energy for future use instead of selling it for short-term gains. Consequently, this leads to relatively poor annual performance. Our proposed L2 regularization addresses this issue by introducing a penalty for unused battery capacity at the end of the day. Additionally, we compared our approach against other benchmark methods, all of which attempt to mitigate the end-of-window effect through alternative means rather than modifying the objective function, but our method consistently outperforms of them.
\section{\label{cha:conc}Conclusions and Future Work}

In this section, we first summarize the key contributions and findings of our work, followed by a discussion of potential ideas for future work.

\subsection{Summary of contributions}

The first contribution of our work is the development of a methodology for determining a price, or more specifically, a price range, for renting battery capacity to energy communities. We show that the proposed stacked revenue model can be advantageous for both the community and the company across multiple tariff types. Moreover, we extend this methodology to also find the optimal battery capacity and generation scale for the community, such that the total cost is minimized. 

We also compare the greedy model against the linear model across multiple types of tariffs to study how the battery should be controlled. Two regularization costs were added to the linear model in an effort to steer the schedule toward more desirable solutions. The first regularization cost, L1, was successful in reducing the number of cycles performed by the battery without increasing the final cost. The L2 regularization exceeded expectations and significantly improved the performance of the linear model. The performance of the combined L1+L2 regularizations was compared against a number of existing state-of-the-art methods for extending LPs to deal with end-of-window effects, and showed to outperform them, for a range of parameters and scenarios.




In summary, our results show that it is feasible for battery owners operating a stacked revenue model to replace some of the profits from the day-ahead market with the newly proposed revenue source of renting part of the battery to energy communities. The proposed methodology is profitable across a wide range of tariffs and can achieve equal or better earnings than the day-ahead prices in the Netherlands using the historical prices from 2023, up to certain capacities. We have shown how this methodology can be extended to calculate the optimal rental sizes. In the case of a flat import tariff of forty cents per kWh, a community of 200 households equipped with a 330 kW wind turbine behind a neighborhood transformer would obtain maximum savings for a capacity of 240 kWh and 280 kWh, depending on whether the community is paid for their exported energy or not. The annual savings associated with these capacities, after subtracting the battery rental costs, are €6,518 and €12,874 respectively. 



\subsection{Future Work}

We see a number of areas as promising for future research. 
First, in this work, we have focused on the flexibility for energy communities and the day-ahead market as potential revenue streams, these being the most relevant and quantifiable for our case study from the Netherlands. Yet our methodology could be extended to include other revenue streams, such as from providing balancing or ancillary services to system operators, in the case of the Netherlands, through the GOPACS platform. This approach could also include considerations such as how volatile/uncertain income is from some income streams (e.g. participation in imbalance markets) vs. income from energy communities, which, while lower, is more stable, leading to a potentially more sustainable business model.

Second, the renewable generation sources considered in this work are wind turbines, as these are often the most widely used form of renewable generation in northwestern Europe. However, we can consider communities that include or rely on solar generation, and study the optimal mix and sizing of renewable generation resources and local battery capacity. This could be particularly relevant if we want to apply our methods to other geographical regions, including countries in the global south, where solar generation is more common. More complex community models can include other distributed forms of energy demand and storage, such as electric vehicle battery charging~\cite{JAIR_EVcharging} or a demand-side response scheme~\cite{antonopoulos_EGAI}. Moreover, while this paper focused on financial metrics such as costs and profits, some energy communities may be more interested in maximizing non-financial objectives, such as the degree of energy autarky (i.e., independence from the central power grid) or self-consumption of locally generated renewable energy.  

In terms of methodology, while our method develops a method to compute the minimum and maximum ranges for battery capacities and prices for which battery-as-a-service (BaaS) models to communities which economic sense to all parties, we have not considered determining how the exact price for a particular contract would be determined. For this, methods in AI-mediated automated negotiation (similar to \cite{etukudor2020automated,Zhang_al}) or coalition formation (as in~\cite{Cremers_Shapley_APEN,groupbuying_TSG}) can be developed and applied, tailored for battery participation.

Finally, a big topic is the integration between battery optimization and control methods with dynamic predictions and forecasts (for both renewable generation and demand) produced by machine learning models.
Incorporating the uncertainties in the forecasts inside linear optimization models for battery control, in a way that maximizes the expected returns, is an open challenge -- but an increasingly important one in a future with high intermittent renewable generation.



\appendix
\section{Computation of Renewable Generation from Wind Speed Data \label{appendix:wind_generation}}

The wind generation curve can be estimated from wind speeds using a methodology similar to Fruh~\cite{Fruh2013Feb, Fruh2015}, Andoni et al.~\cite{Andoni2017Sep,andoni_IEEEAccess}, and Norbu et al.~\cite{Norbu2021Apr}.

Real wind speed data collected by the Royal Netherlands Meteorological Institute (KNMI) was used for the calculation. KNMI provides climate data averaged over several decades for 46 weather stations, as well as the provincial averages of the 12 Dutch provinces. For this study, data from North Holland was used, as this is where all of GIGA's assets are located, and the community should be situated around these assets. 

\begin{figure}
\begin{minipage}{.5\textwidth}
    \centering
    \includegraphics{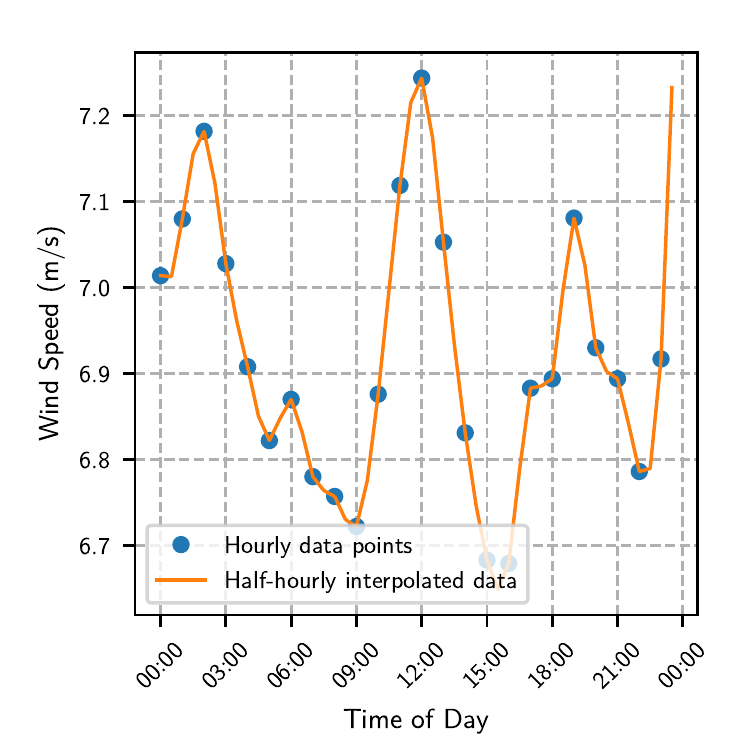}
     \parbox{0.9\textwidth}{\caption{Historical wind speed data from North Holland recorded by KNMI, with interpolated values for missing timesteps.}\label{fig:wind-speeds}}
\end{minipage}%
\begin{minipage}{.5\textwidth}
    \centering
    \includegraphics{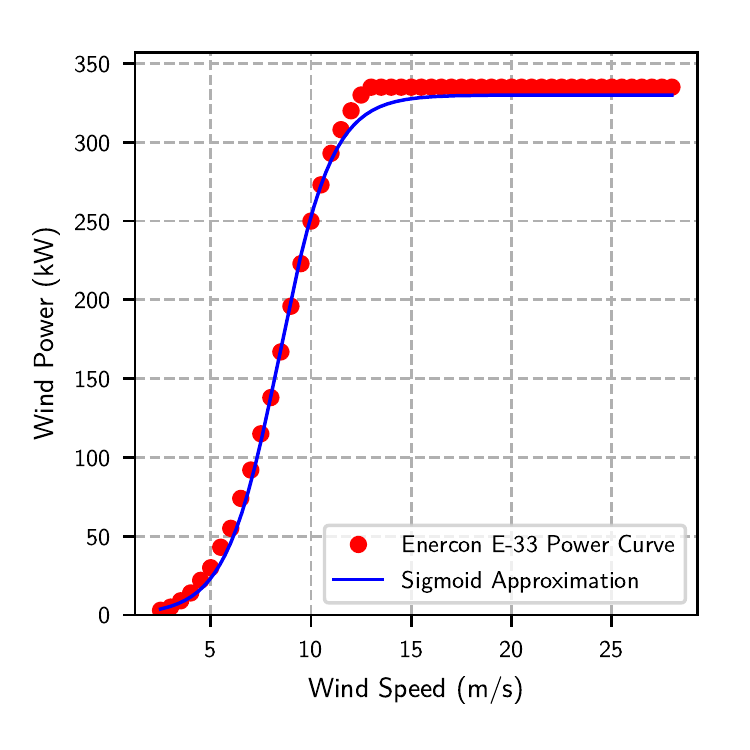}{\caption{Comparison of the power curve of the Enercon E-33 wind turbine \cite{EnerconE33} with the sigmoid approximation described by Eq. \ref{eq:methodology-sigmoid}.} \label{fig:methodology-enercon-power-curve}}
\end{minipage}
\end{figure}

The data was retrieved from TU Delft's Meteorological data portal\footnote{\url{https://www.tudelft.nl/en/ewi/over-de-faculteit/afdelingen/electrical-sustainable-energy/photovoltaic-materials-and-devices/dutch-pv-portal/meteorological-data}}, and it consists of average wind speeds for a year with hourly time steps, measured at a height of 10m above the ground. Any missing data is computed using a double spline interpolation function. The historical and interpolated wind speeds are shown in Figure \ref{fig:wind-speeds}. 

Wind turbines generally operate at greater altitudes, so a logarithmic shear profile is used to extrapolate wind speeds to the required height. The calculation is expressed as:

$$
 u_h = u_a \frac{log z_h/z_0}{log z_a/z_0} 
$$
where $u_h$ is the wind speed at the wind turbine hub height $z_h = 50 \, \text{m}$, $u_a$ is the wind speed measured at the anemometer height $z_a = 10 \, \text{m}$, and $z_0 = 0.03 \, \text{m}$ is the surface roughness of grass - a typical environment around weather stations. This approach follows the methodology used in \cite{Fruh2013Feb, Fruh2015, Andoni2017Sep, Norbu2021Apr}. 

The final power output can be approximated using the power curve provided by the manufacturer, as seen in Figure \ref{fig:methodology-enercon-power-curve}. For this study, we will use the power curve of the Enercon E-33 wind turbine \cite{EnerconE33}, which can be approximated by a sigmoid function with parameters $ a = 0.7526 s/m$ and $b = 8.424 m /s$:

\begin{equation}
   f(x,a,b) = \frac{1}{1 + e ^ {-a(x-b)}} \label{eq:methodology-sigmoid}
\end{equation}

The yearly cost of the wind turbine was calculated and added to the community's final bill. Without including this cost, energy generation would appear free, making a larger turbine always advantageous, which is unrealistic. Incorporating this cost allows for optimizing the scale of generation by considering the turbine's expense. The installation cost for wind turbines has steadily declined in recent years. For our study, we used the average weighted price from 2022 of €1200 per kW of installed capacity \cite{StatistaWindInstalation}. The wind turbine's cost is amortized over 20 years, resulting in a yearly cost of €19,800. One cost not considered in this paper is the maintenance cost of the wind turbine. This cost is more challenging to compute as it depends on multiple factors, such as the age and capacity of the turbine and the type of technology used. Therefore, this cost was not included, but readers should be aware of these costs, usually referred to as operations and maintenance (O\&M) costs \cite{OMcosts}.

Finally, we need to ensure that the sizes of local renewable generation capacity and demand are aligned at realistic levels relative to each other. This is achieved through a \emph{wind generation coefficient} $R_{gen}$ for the installed renewable generation. As discussed, the renewable scaling coefficient is defined as the ratio between the yearly demand of the community and the total generation of the community wind turbine for that year.

For the real demand data, type of wind turbine, and wind speed data considered in this study (i.e. demand from 200 households from the Thames Valley Vision trial, the Enercon E-33 wind turbine \cite{EnerconE33}, and wind speeds from KNMI), $R_{gen}\approx1.2$. Intuitively, the total generation from the community wind turbine, computed over all hours in 2023, is 20\% above the total community energy demand for that year. 

In summary, Table~\ref{tabe:appendix_param_values} provides the values of all the parameters used in this appendix.
\begin{table}[h]
\centering
\caption{Summary of parameter values used in wind power computations.}
\label{tabe:appendix_param_values}
\begin{tabular}{|l l|l l|}
\hline
\textbf{Parameter} & \textbf{Value} & \textbf{Parameter} & \textbf{Value} \\
\hline
$z_h$ & 50 m (hub height) & $z_a$ & 10 m (anemometer height) \\
$z_0$ & 0.03 m (surface roughness) & $R_{\text{gen}}$ & 1.2 \\
$a$ & 0.7526 s/m (sigmoid slope) & $b$ & 8.424 m/s (sigmoid midpoint) \\
Turbine cost & €1200 per kW & Capacity & 330 kW \\
Amortization period & 20 years & Yearly cost & €19,800 \\
\hline
\end{tabular}
\end{table}

\section{\label{appendix:battery_degradatioj_model}Battery Degradation Model} \label{app_subsec:battery_degradation}

In this section, we describe the methodology for computing the battery's depreciation factor (DF). This is included for completeness, and it was first presented in prior work of (some of) the authors of this paper \cite{Norbu2021Apr}. Frequent charging and discharging of the battery, especially when the depth of discharge is deep, can accelerate degradation significantly. Manufacturers of batteries often specify a battery cycle life on a provided datasheet. In this study, we use the lithium-ion battery cycle life data specified by \cite{xu2018modeling}. This cycle life specifies the expected number of charge/discharge cycles per depth of discharge (DoD) that can be performed before the performance drops below operable levels. A shortened lifetime of the battery can have an impact on the depreciation cost, hence, the inclusion of the depreciation factor provides a more realistic estimate of the community cost.

The total depreciation factor of the battery can be found by computing the depreciation factors of regular and irregular cycles. A regular cycle starts its discharging phase from an SoC level of 100\% and is charged back to 100\% SoC level. The depreciation factor of regular cycles is then computed as follows:
\begin{equation}
    \text{DF}^{\text{regular}} = \sum_{DoD=0\%}^{100\%} \frac{n^{\text{DoD,regular}}_{cycles}}{N^{\text{DoD,max}}_{cycles}}
\end{equation}
where $n^{\text{DoD, regular}}_{cycles}$ is the number of regular cycles (starting from 100\% SoC) with the DoD value, and $N^{\text{DoD,max}}_{cycles}$ is the lifetime of the battery provided by the battery manufacturer in terms of the number of regular cycles the battery can go through with the DoD value.

However, not all cycles can start from 100\% SoC level. Cycles that have a starting point other than 100\% SoC are called irregular cycles. Regular and irregular cycles can have the same DoD, e.g., a regular cycle that starts discharging at 100\% SoC level until it reaches 60\% then charged back to 100\% and an irregular cycle with 80\% starting SoC discharging until 40\% SoC level and charged back to 80\% SoC, both cycles have the same DoD of 40\%. Yet, regular and irregular cycles have different impacts on the depreciation factor, and hence, they are computed separately. Furthermore, an irregular cycle can either be a full or half cycle. A full cycle consists of both discharging and charging phases, whereas a half cycle is only one of the charging/discharging phases. The rain-flow cycle counting algorithm \cite{ke2015control, Norbu2021Apr} is used to classify and count the cycles as regular or irregular, as well as full or half. 

For all the irregular cycles $L$ during the evaluation period, we compute the depreciation factor as follows:
\begin{equation}
    \text{DF}^{\text{irregular}} = \sum_{l \in L} \text{n}_l \times \left| \frac{1}{N^{\text{DoD}^{eq}(SoC^{l,Start}_{\%}),max}_{cycles}} - \frac{1}{N^{\text{DoD}^{eq}(SoC^{l,End}_{\%}),max}_{cycles}}\right|
\end{equation}
where $SoC^{l,Start}_{\%}$ and $SoC^{l,End}_{\%}$ are the starting and ending state of charges, respectively, for the irregular cycle $l$, and $\text{n}_l$ represents the type of the cycle ($1$ for full cycle, and $\frac{1}{2}$ for half cycle). Then, $N^{\text{DoD}^{eq}(SoC^{l,Start}_{\%}),max}_{cycles}$ corresponds to the maximum number of cycles the battery can perform for $\text{DoD}^{eq}(SoC^{l,Start}_{\%})$, i.e. a depth of discharge equivalent to a cycle starting at 100\% $SoC_{\%}$ and ending at $SoC^{l,Start}_{\%}$. This is computed as follows:
\begin{equation}
    \text{DoD}^{eq}(SoC^{l,Start}_{\%}) = 100 - \left(\frac{SoC^{l,Start}_{\%}}{SoC^{\text{max}}_{\%}} \times 100 \right)
\end{equation}
We compute $N^{\text{DoD}^{eq}(SoC^{l,End}_{\%}),max}_{cycles}$ using a similar notion.

Finally, given the depreciation factors of regular and irregular cycles, the total depreciation factor of the battery is the sum of the two components: 
\begin{equation}
    \text{DF} = \text{DF}^{\text{regular}} + \text{DF}^{\text{irregular}}
\end{equation}

\section{Comparison Between the MILP and LP Models}
\label{appendix-comparison-milp-lp} 

\begin{figure}
    \centering
    \includegraphics[]{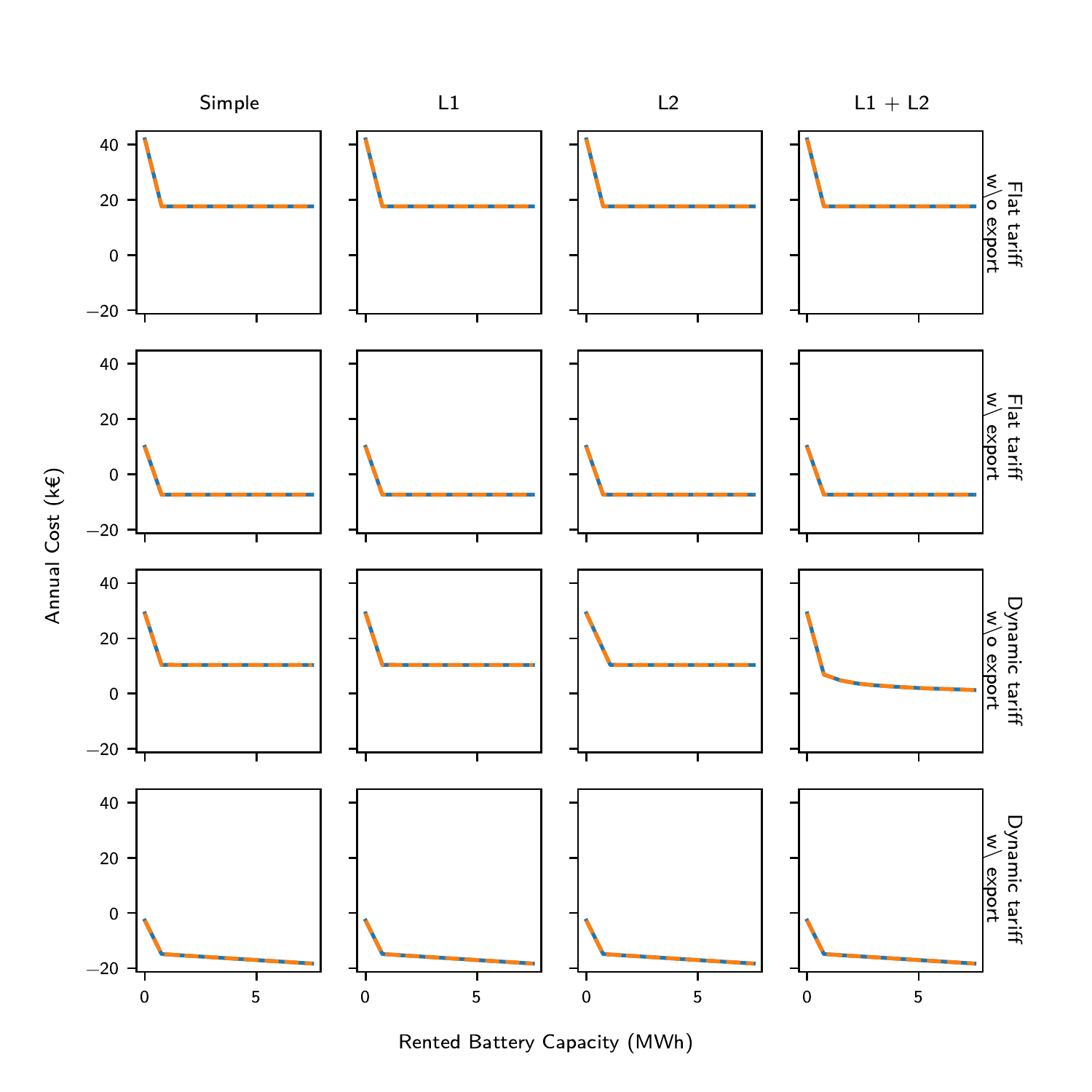}
    \caption{Daily comparison of LP (blue) and MILP (orange) models, initialized with zero state of charge (SoC) at the start of each day. Results demonstrate identical daily optimal solutions across all scenarios. Each column represents a different regularization setup: No regularization (first), L1 only (second), L2 only (third), and both L1 and L2 (fourth). Rows correspond to tariff cases: flat tariff without export (first row), flat tariff with export (second row), dynamic tariff without export (third row), and dynamic tariff with export (fourth row).}
    \label{fig:comparizon_MILP_LP_zero}
\end{figure}

In this Appendix, for completeness, we provide a comparison of the linear programming (LP) model and mixed integer linear programming (MILP) model for this problem, essentially showing their equivalence in terms of the optimal costs returned by the two models. We include this section as an appendix, because all result graphs are essentially overlapping, thus of limited interest for our main findings.

As discussed in Sections~\ref{sec:battery_control} and~\ref{sec:meth:linear-programing-simple}, in addition to the constraints of the LP model (shown in \hyperref[sec:meth:linear-programing-simple]{Model 1}), the MILP model adds additional constraints (shown in \hyperref[alg:methodology-mixed-linear-model]{Model 3}) to enforce mutually exclusive operations, i.e. charging versus discharging or importing versus exporting. To eliminate simultaneous operations, the MILP model introduces binary variables for battery and grid operations, ensuring mutually exclusive actions. Without these constraints, it is conceptually possible that simultaneous operations may occur, potentially distorting the objective function. For instance, Pozo '22~\cite{POZO2022108565} showed that LP formulations could overestimate objectives by up to 15\% in set-point tracking (SPT) problems. Yet, as the setting studied here includes only one battery and one renewable source, such solutions, while theoretically possible, are never in the optimal solution set, making these additional constraints unnecessary to include explicitly. 

In most cases, MILP solving times are comparable to LP ones, with both typically completing within seconds. However, there are a small number of instances (a few percentage points of the total) where the MILP formulations require significantly longer solving times. We attribute this effect (also shown in the work of Pozo~\cite{POZO2022108565}), to the specific optimization paths taken by MILP solvers when adding binary exclusion constraints. Because, for our set-up, there are no differences in the solutions found by the two solvers (as shown in this Appendix), we consistently report the LP results in the main body of the paper.

\begin{model}
\caption{Mixed Integer Linear Programming Model for Battery Control (Extension of \hyperref[sec:meth:linear-programing-simple]{Model 1})}\label{alg:methodology-mixed-linear-model}
\begin{algorithmic}
    \State \textbf{Additional Constraints:}
    \State $p_{i}^{charge} \leq p^{max} \cdot (1 - x_i)$
    \State $p_{i}^{discharge} \leq p^{max} \cdot x_i$
    \State $e^b_{i} \leq e^{max} \cdot (1 - y_i)$
    \State $e^s_{i} \leq e^{max} \cdot y_i$
    
    \hfil
    
    \State \textbf{Binary Decision variables:}
    \State $x_i \in \{0, 1\}$ \Comment{Binary variable indicating if the battery is discharging at time step $i$}
    \State $y_i \in \{0, 1\}$ \Comment{Binary variable indicating if the grid is exporting energy at time step $i$}
\end{algorithmic}
\end{model} 

In order to validate that indeed the two approaches output solutions with the same costs, we re-ran experiments with the battery initialized to zero at the start of each day. Under these conditions, LP and MILP produced identical results across all scenarios (Figure \ref{fig:comparizon_MILP_LP_zero}), confirming their equivalence for daily optima. These additional experiments clearly show that additional binary charge/discharge mutual exclusion constraints are unnecessary in our setting, as the LP model consistently identifies correct solutions.

\vspace{-0.1cm}
\section*{Acknowledgements}
\vspace{-0.25cm}
Valentin Robu and Tudor Octavian Pocola acknowledge the support of the project ``TESTBED2: Testing and Evaluating Sophisticated information and communication Technologies for enaBling scalablE smart griD Deployment'', funded by the European Union under the Horizon2020 Marie Skłodowska-Curie Actions (MSCA) [Grant agreement number: 872172]. Benoit Couraud, Merlinda Andoni and David Flynn acknowledge the support of the InnovateUK Responsive Flexibility (ReFLEX) project [ref: 104780]. Merlinda Andoni and David Flynn also acknowledge the support of the UK Engineering and Physical Science Research Council in the project ``DecarbonISation PAThways for Cooling and Heating (DISPATCH)"" [grant: EP/V042955/1]. Sonam Norbu acknowledges the support of the InnovateUK Knowledge Transfer Partnerships (KTP) Project based at The Crichton Trust in Dumfries [KTP-13052]. H. Vincent Poor acknowledges the support of the U.S. National Science Foundation under Grant ECCS-2039716 and a grant from the C3.ai Digital Transformation Institute.
 \bibliographystyle{elsarticle-num} 
 \bibliography{mybib}

\end{document}